\newcommand\vx{\vec{x}}
\newcommand\vr{\vec{r}}
\newcommand\vk{\vec{k}}
\newcommand\vs{\vec{s}}
\newcommand\hk{\hat{k}}

\newcommand\Ng{N_{\rm g}}

\newcommand\hr{\hat{r}}

\newcommand\rs{r_{\rm s}}

\newcommand\omegam{\Omega_{\rm m}}
\newcommand\omegab{\Omega_{\rm b}}

\newcommand\oO{\mathcal{O}}

\newcommand\Vmax{V_{\rm max}}
\newcommand\Rmax{R_{\rm max}}
\newcommand{\specialcell}[2][l]{%
  \begin{tabular}[#1]{@{}l@{}}#2\end{tabular}}

\newcommand\fb{f_{\rm b}}

\newcommand\Mpch{{\rm\; Mpc}/h}

\documentclass[useAMS,usenatbib]{mn2e}

\usepackage{lineno}
\usepackage{placeins}

\usepackage{graphicx} 
\usepackage{multirow}
\usepackage{array}

\usepackage{amsmath,amssymb}

\usepackage[T1]{fontenc}
\usepackage[latin9]{inputenc}
\usepackage{float}
\usepackage{graphicx}
\usepackage{esint}

\makeatletter
\DeclareFontEncoding{LGR}{}{}
\DeclareTextSymbol{\~}{LGR}{126}
\begin{document}

\title[DR12 CMASS large-scale 3PCF]{The large-scale 3-point correlation function of the SDSS BOSS DR12 CMASS galaxies}

\author{\makeatauthor}

\author[Slepian et al.]{Zachary Slepian$^{1}$\thanks{zslepian@cfa.harvard.edu},
Daniel J. Eisenstein$^{1}$\thanks{deisenstein@cfa.harvard.edu},
Florian Beutler$^{2}$,
Antonio J. Cuesta$^{3}$, \and
Jian Ge$^{4}$,
H\'ector Gil-Mar\'in$^{5,6,7}$,
Shirley Ho$^8$,
Francisco-Shu Kitaura$^9$,\and
Cameron K. McBride$^{1}$, 
Robert C. Nichol$^{7}$,
Will J. Percival$^{7}$,\and 
Sergio Rodr\'iguez-Torres$^{10,11,12}$,
Ashley J. Ross$^{13}$,
Rom\'an Scoccimarro$^{14}$, Hee-Jong Seo$^{15}$,\and
Jeremy Tinker$^{14}$, %this is in the wrong place! should go after r. 
Rita Tojeiro$^{16}$,
\& Mariana Vargas-Maga\~na$^{17}$\\
$^{1}$ Harvard-Smithsonian Center for Astrophysics, 60 Garden Street, Cambridge, MA 02138, USA\\%ZS, DE, CM
$^2$ Lawrence Berkeley National Lab, 1 Cyclotron Rd, Berkeley CA 94720, USA\\%FB
$^3$ Institut de Ci{\`e}ncies del Cosmos (ICCUB), Universitat de Barcelona (IEEC-UB), Mart{\'\i} i Franqu{\`e}s 1, E08028 Barcelona, Spain\\%AJC
$^4$ Astronomy Department, University of Florida, 211 Bryant Space Science Center, Gainesville, FL 32611, USA\\%Jian Ge
$^5$ Sorbonne Universit\'es, Institut Lagrange de Paris (ILP), 98 bis Boulevard Arago, 75014 Paris, France\\%HGM
$^6$ Laboratoire de Physique Nucl\'eaire et de Hautes Energies, Universit\'e Pierre et Marie Curie, 4 Place Jussieu, 75005 Paris, France\\%HGM
$^7$ Institute of Cosmology \& Gravitation, University of Portsmouth, Dennis Sciama Building, Portsmouth PO1 3FX, UK\\%HGM, WJP, BN
$^8$ Bruce and Astrid McWilliams Center for Cosmology,
Department of Physics, Carnegie Mellon University,\\%SH
5000 Forbes Ave, Pittsburgh, PA 15213, USA\\
$^9$ Leibniz-Institut f\"{u}r Astrophysik Potsdam (AIP), An der Sternwarte 16, D-14482 Potsdam, Germany\\%Francisco-Shu Kitaura (also Chia-Hsun Chuang)
$^{10}$ Instituto de F\'isica Te\'orica, (UAM/CSIC), Universidad Aut\'onoma de Madrid, Cantoblanco, E-28049 Madrid, Spain\\%SRT
$^{11}$ Campus of International Excellence UAM+CSIC, Cantoblanco, E-28049 Madrid, Spain\\%SRT
$^{12}$ Departamento de F\'isica Te\'orica M8, Universidad Autonoma de Madrid (UAM), Cantoblanco, E-28049 Madrid, Spain\\%SRT
$^{13}$ Center for Cosmology and Astroparticle Physics, Department of Physics, The Ohio State University, OH 43210, USA\\%AR
$^{14}$ Center for Cosmology and Particle Physics, New York University, 4 Washington Place, NY 1003, New York, USA\\%RS,  JT
$^{15}$ Department of Physics and Astronomy, Ohio University, Clippinger Labs, Athens, OH 45701\\%HJS
$^{16}$ University of St Andrews, North Haugh, St Andrews Fife, KY16 9SS, UK\\%Rita Tojeiro 
$^{17}$ Instituto de F\'isica, Universidad Nacional Aut\'onoma de M\'exico, Apdo. Postal 20-364, M\'exico}%Mariana Vargas-Magana.

\maketitle

\begin{abstract}
We report a measurement of the large-scale 3-point correlation function of galaxies using the largest dataset for this purpose to date, $777,202$ Luminous Red Galaxies in the Sloan Digital Sky Survey Baryon Acoustic Oscillation Spectroscopic Survey (SDSS BOSS) DR12 CMASS sample. This work exploits the novel algorithm of Slepian \& Eisenstein (2015b) to compute the multipole moments of the 3PCF in $\oO(N^2)$ time, with $N$ the number of galaxies. Leading-order perturbation theory models the data well in a compressed basis where one triangle side is integrated out. We also present an accurate and computationally efficient means of estimating the covariance matrix. With these techniques the redshift-space linear and non-linear bias are measured, with $2.6\%$ precision on the former if $\sigma_8$ is fixed. The data also indicates a $2.8\sigma$ preference for the BAO, confirming the presence of BAO in the 3-point function.
\end{abstract}

\section{Introduction}
The spatial distribution of galaxies on large scales contains a wealth of information on both the Universe's contents and the laws under which these contents evolved.  The 3-point correlation function (3PCF) is an important tool for quantifying this distribution (Groth \& Peebles 1977; Peebles 1980; Bernardeau et al. 2002). It measures the excess probability over random of finding three galaxies whose separations form a given triangle. In the consensus picture of structure formation, density perturbations were set up at the end of inflation as a nearly Gaussian random field (Starobinsky 1980; Bardeen, Steinhardt \& Turner 1983).  These primordial fluctuations are thus expected to be nearly fully characterised by their 2-point correlation function (2PCF), with a negligible 3PCF (Planck Paper XVII, 2015).  However, subsequent evolution generates a 3PCF as well as imprinting additional information in the 2PCF.  Further, the small level of primordial non-Gaussianity (PNG) generically predicted at the end of inflation produces its own unique signature in the 3PCF (Desjacqes \& Seljak 2008, for a review; Sefusatti \& Komatsu 2007 for forecasts; for recent observational work using large-scale structure, see Xia et al. 2010, 2011; Ross et al. 2013; Giannantonio et al. 2014; for Cosmic Microwave Background constraints, see Planck Paper XVII, 2015).

This picture applies within the context of perturbation theory (PT) for the cosmic density field, i.e for small perturbations around a homogeneous background.  The galaxies we observe are tracers of this density field, but do so with some bias because galaxy formation is a complicated process (Kaiser 1984; Dekel \& Rees 1987).  In addition to information on the initial conditions and on high redshift physics, the 2PCF and 3PCF therefore illuminate galaxy formation, revealing how faithfully galaxies trace the underlying matter density.  

Since the 1970s, both 3PCF and 2PCF have been measured in galaxy redshift surveys of increasingly large volume and number of objects.  In the last decade a particular focus of 2PCF measurements has been to use the Baryon Acoustic Oscillation (BAO) method, which exploits a sharp bump in the 2PCF, imprinted by sound waves in the early Universe, to constrain the Universe's expansion history and thus the cosmological parameters (Eisenstein, Hu \& Tegmark 1998; Blake \& Glazebrook 2003; Hu \& Haiman 2003; Linder 2003; Seo \& Eisenstein 2003). The 3PCF is intrinsically weaker than the 2PCF, and consequently has not been a powerful additional lever on the BAO scale. Moreover, BAO analyses now routinely use reconstruction, a technique to move galaxies backwards along their inferred velocity vectors to try to undo some of the low-redshift non-linear evolution. Reconstruction reduces the amplitude of the 3PCF further and puts some of the distance information in the 3PCF back into the 2PCF (Eisenstein et al. 2007; Padmanabhan, White \& Cohn 2009; Burden et al. 2014). Schmittfull et al. (2015) explicitly shows how 3- and 4-point statistics of the unreconstructed density field enter the reconstructed 2PCF.

Measurements of the 3PCF have further been complicated by its scaling with the number of objects. In contrast to the 2PCF, which requires pair counting and hence scales as $N(n\Vmax)$, the 3PCF involves triplets of galaxies and so scales as $N(n\Vmax)^2$, where $N$ and $n$ are the number and number density of galaxies and $\Vmax$  is the volume within a sphere of radius $\Rmax$, the maximal scale at which the correlations are measured.  While $n\Vmax$ is much smaller than $N$ for large-volume surveys, this scaling has nonetheless mostly inhibited exploitation of the 3PCF on the acoustic scale of $\rs \sim 100\Mpch$.  It is not only measuring the 3PCF of the data that is time-consuming; in most methods, the number of points must also include a set of random points, much larger than the actual galaxy set, in order to model the survey geometry. Further, computing covariance matrices to fit parameters requires using potentially thousands of mock catalogs generated by N-body simulations combined with some prescription for matching galaxies to dark matter halos.  There is also a problem of visualization: the 3PCF depends on three independent variables (three triangle sides, or two triangle sides and an angle), so it has not been clear how to show the results of an analysis. In practice what is often done is to show the angle dependence of the amplitude for certain fixed triangle side lengths or ratios.

For these reasons, 3PCF measurements have been limited in ability to exploit the full data sets used by the 2PCF. They often have focused on only certain triangle configurations (two-to-one, isosceles) and worked on scales smaller than the BAO scale. Even so these measurements have been computationally challenging.  

Historically the 3PCF traces back to the pioneering work of Groth \& Peebles (1977); our list of works here focuses on the recent and is not intended to be exhaustive.\footnote{Different authors use different parametrizations; to standardize we quote the largest triangle side possible in each work.} The largest sample used for the 3PCF to date is 220,000 galaxies from SDSS DR6 in McBride et al. (2011a,b), with 2-to-1 triangles up to 27 $\Mpch$. There have been a number of other works also using SDSS data. Nichol et al. (2006) use 36,738 LRGs for 2-to-1, 3-to-1, and 4-to-1 triangles on scales up to $\sim 40\Mpch$; Kulkarni et al. (2007) use 50,967 SDSS DR3 LRGs for isosceles, 2-to-1, and 3-to-1 triangles on scales up to $30\Mpch$; Mar\'in (2011) uses 105,831 Luminous Red Galaxies (LRGs) from SDSS DR7 for 2-to-1 triangles on scales up to $90\Mpch$; Guo et al. (2014) use DR7 LRGs to measure isosceles, 2-to-1, and 3-to-1 configurations up to $40\Mpch$. Guo et al. (2015) use LRGs in the DR11 CMASS sample to measure 2-to-1 configurations up to $40\Mpch$.  

Data from the Two-degree-Field Galaxy Redshift Survey (2dFGRS) has also been used for 3PCF measurements: Jing \& B\"{o}rner (2004) and Wang et al. (2004) use respectively $\sim 60,000$ and $\sim 250,000$ galaxies from this survey on scales up to respectively $40\Mpch$ and $\sim 20\Mpch$. The 3PCF has also been measured using the WiggleZ Dark Energy Survey: Mar\'in et al. (2013) use  isosceles, 2-to-1, and 3-to-1 triangles for scales up to $90\Mpch$.  

By concentrating only on particular triangle configurations, these measurements leave information in the 3PCF unexploited. Importantly, they also do not probe the BAO scale, which in principle offers cosmological distance information.

The only work to probe the BAO scale in the 3PCF thus far has been Gazta\~naga et al. (2009) using a sample of $\sim 40,000$ LRGs from SDSS DR7. They find a $2-3\sigma$ detection of the BAO using all opening angles of a single triangle configuration with side lengths $r_1=33\Mpch$ and $r_2=88\Mpch$. They did not fit for the BAO scale to extract distance information.

Similar datasets to those discussed above have been used for measuring the bispectrum, the Fourier space analog of the 3PCF.  While in principle the signal-to-noise is the same for the bispectrum and the 3PCF, in practice we believe Fourier space has several disadvantages relative to configuration space in this context. First, the bispectrum requires gridding the galaxies to a regular mesh to allow an $\Ng\log \Ng$ Fourier Transform (FT) to be computed, with $\Ng$ the number of grid cells. Consequently, it only samples a discrete set of wavenumbers, so some spatial information is lost. Second, in practice there is a truncation of the transform at high wavenumber, set by the Nyquist frequency given by the grid spacing. Third, edge correction is not straightforward in the bispectrum, as even simple survey boundaries in real space lead to complicated, infinite-support ringing in Fourier space.  Fourth, the BAO features in the 3PCF are localized but in the bispectrum are dispersed, much as the sharp bump in the 2PCF corresponds to extended wiggles in the power spectrum. This localization is likely to make it easier in practice to fit for BAO information in the 3PCF than in the bispectrum.

Nonetheless, a major advantage of the bispectrum is that it is obvious how to estimate it quickly. Even the most naive approach only requires comparison of pairs of wavevectors rather than galaxy triplets. Additional speed can come from using templates to directly extract bias parameters within a given bispectrum model if the angular and wave-vector-length dependences are separable (Schmittfull, Baldauf \& Seljak 2015); there are also accelerations available in the absence of separability provided the bispectrum model can be expanded in a finite number of separable basis functions (Schmittfull, Regan \& Shellard 2013). Further, it has long been known that the opening-angle-averaged bispectrum can be computed using FTs (Scoccimarro 2000), a result recently extended to the full, unaveraged bispectrum (Scoccimarro 2015). Previous work on the bispectrum has been Feldman et al. (2001), Scoccimarro et al. (2001), and Verde et al. (2002), with the most recent measurement, using  690,827 LRGs from the SDSS DR11 CMASS sample, by Gil-Mar\'in et al. (2015).  This measurement uses isosceles and 2-to-1 triangles in Fourier space up to scales of $k=0.03\;h/{\rm Mpc}\sim 30\Mpch$.  Gil-Mar\'in et al. (2015) use only diagonal elements of the covariance matrix for parameter fitting since the large number of matrix elements relative to the number of mock catalogs available means the full covariance cannot be accurately computed empirically.

In this paper, we present the first measurements of the large-scale 3PCF for the CMASS sample of $777,202$ LRGs within SDSS DR12, working on scales from $75-180\Mpch$.  We use the novel algorithm of Slepian \& Eisenstein (2015b), hereafter SE15b, which computes the 3PCF's multipole moments in $\oO(N(n\Vmax))$ time using spherical harmonic decompositions; the covariance matrix also turns out to be tractable in the multipole basis (SE15b). The main outcomes of this work are:

\indent{\bf 1)} A highly accurate covariance matrix derived analytically but matching mock catalogs well.\\
\indent {\bf 2)} Demonstrating that perturbation theory is in agreement with the data on these scales.\\
\indent {\bf 3)} Measurement of the redshift-space linear and non-linear bias with $2.6\%$ precision on the former.\\
\indent{\bf 4)} A $2.8\sigma$ detection of the BAO feature in the 3PCF.\\

The paper is laid out as follows. In \S\ref{sec:drm} we briefly introduce the data set, the random catalogs used for edge correction, and the mock catalogs used to verify our analysis pipeline and covariance matrix. \S\ref{sec:basis} presents our multipole basis for the 3PCF and reviews the algorithm used to do the measurement. \S\ref{sec:compression} recapitulates a compression scheme developed in previous work that we apply to the 3PCF here. \S\ref{sec:model} details our perturbation theory model, while \S\ref{sec:covar} describes our approach to obtaining the covariance matrix. \S\ref{sec:results} shows the results of our analysis on both mock catalogs and the CMASS data. \S\ref{sec:conclusions} concludes.

\section{Data, Randoms, and Mocks}
\label{sec:drm}
Here we summarize the data set, random catalogs, and mock catalogs used. The data are $777,202$ LRGs within the CMASS sample of DR12 (Alam et al. 2015) of the Baryon Oscillation Spectroscopic Survey (BOSS; Eisenstein et al. 2011; Dawson et al. 2013) in the redshift range $0.43$ to $0.7$, with $568,776$ in the North Galactic Cap over $6,934$ square degrees and $208,426$ in the South Galactic Cap over $2,560$ square degrees, for a total of $9,493$ square degrees (Reid et al. 2016, Table 2). The sample is color-selected to have roughly constant stellar mass and observations were completed in 2014.  Target selection and the algorithms used for creating the galaxy catalogues and the associated random catalogues are presented in Reid et al. (2016), with the treatment for observational systematic biases described in Ross et al. (2012) and Ross et al. (2015). The 2PCF of this dataset is used to constrain the cosmic distance scale via the BAO method in Cuesta et al. (2015).

In total, the SDSS (York et al. 2000), comprising SDSS I, II (Abazajian et al. 2009) and SDSS III (Eisenstein et al. 2011) imaged more than one-third of the sky (14,555 square degrees) with a drift-scanning mosaic CCD camera (Gunn et al. 1998) in five photometric bandpasses (Fukugita et al. 1996; Smith et al. 2002; Doi et al. 2010) on the 2.5-m Sloan Telescope (Gunn et al. 2006) at Apache Point Observatory in New Mexico.  The data underwent astrometric calibration (Pier et al. 2003), photometric reduction (Lupton et al. 2001), photometric calibration (Padmanabhan et al. 2008), and reprocessing for Data Release 8 (Aihara et al. 2011). Within BOSS, targets were assigned to tiles with an adaptive algorithm (Blanton et al. 2003), spectra obtained using double-armed spectrographs (Smee et al. 2013), and redshifts derived as described in Bolton et al. (2012).

We also ran our pipeline on 299 mock catalogs developed for DR12 known as the \textsc{MultiDark-Patchy} BOSS DR12 mocks (Kitaura, Yepes \& Prada 2014; Kitaura et al. 2015a,b). These catalogs used second-order Lagrangian perturbation theory (2LPT) combined with a spherical collapse model on small scales (Kitaura \& He{\ss} 2013).  They were calibrated on accurate N-body-based reference catalogs using halo abundance matching to reproduce the number density, clustering bias, selection function, and survey geometry of the BOSS data (Rodr\'iguez-Torres et al. 2015).

\section{Basis and algorithm}
\label{sec:basis}
We parameterize the 3PCF using two triangle sides $r_1$ and $r_2$ and the cosine of the angle between them, $\cos \theta_{12} = \hr_1\cdot\hr_2$.  We then project the angular dependence onto the basis of Legendre polynomials $P_l$ of $\cos\theta_{12}$, so that the 3PCF is written as
\begin{align}
\zeta(r_1, r_2;\hr_1\cdot\hr_2)= \sum_l \zeta_l(r_1,r_2)P_l(\hr_1\cdot\hr_2).
\label{eqn:3pcf_multi}
\end{align}
The multipole coefficients $\zeta_l$ capture the 3PCF's dependence on side lengths at each multipole. We bin in the two side lengths but our measurement uses the exact angle between the triangle sides when projecting onto the multipoles. The Legendre polynomials of even or odd $l$ involve a falling sequence of even or odd powers of $\cos\theta_{12}$, with $l$ giving the maximal power. This expansion was first proposed by Szapudi (2004) but has up to now not been used save for the monopole term. Since $P_0=1$, the monopole corresponds to measuring the angle-averaged 3PCF (Pan \& Szapudi 2005). The motivation for this expansion is that at leading order in perturbation theory, prior to cyclically summing over all triangle vertices, the 3PCF has only $l=0,1,$ and $2$ moments (the perturbation theory (PT) is reviewed in Bernardeau et al. 2002).  Cyclically summing around the triangle vertices then introduces higher moments but they do not contain substantial new information, tending to be dominated by the same population of squeezed triangles. These points are further discussed in Slepian \& Eisenstein (2015a,b), hereafter SE15a and SE15b.

The Legendre polynomials and radial coefficients of equation (\ref{eqn:3pcf_multi}) form a complete, orthogonal basis for any functions depending only on the relative angle $\hr_1\cdot\hr_2$. This basis builds in isotropy, which is in reality not true due to redshift-space distortions (RSD; see Hamilton 1998 for a review).  By using this basis, we spherically average over all orientations to the line of sight, so the effects of RSD get integrated over much as is the case in the spherically averaged, isotropic 2PCF. We discuss RSD further in \S\ref{subsec:rsd}.

We directly measure the 3PCF's multipole coefficients $\zeta_l$ using an algorithm presented in SE15b that substantially accelerates the computation. We
construct the counts over triples of galaxies or randoms as follows.
For each galaxy in the sample successively, we bin the surrounding galaxies
into concentric spherical shells set by our radial binning around that galaxy and store the unit vectors for each of
the separations.  On each spherical shell, we compute the multipole moments
of the unit vectors and then convert these to the spherical harmonic
decomposition $a_{\ell m}$ of the galaxy field.  We then form all of
the cross-powers between the shells as a function of the order $\ell$,
summing over the azimuthal parameter $m$.  We finally sum the resulting
cross-powers over all central galaxies.  The spherical harmonic computation around each galaxy is $\oO(n\Vmax)$ and we must do this around each of the $N$ galaxies in the survey, leading to a 3PCF measurement that scales as $N(n\Vmax)$. Further details such as implementation, tests and runtimes are given in SE15b.

We now briefly recapitulate the edge-correction framework of SE15b.  We measure the projection onto multipoles of the Szapudi-Szalay (1998) estimator for the 3PCF,
\begin{align}
\zeta = \frac{NNN}{RRR}=\sum \zeta_l(r_1,r_2)P_l(\hr_1\cdot\hr_2)
\end{align}
where $N\equiv D-R$, $D$ denotes a data point, and $R$ denotes a random point.  It can be shown that if the $NNN$ and $RRR$ counts are given by the multipole series
\begin{equation}
NNN=\sum_j \mathcal{N}_{j}(r_{1},r_{2})P_{j}(\hat{r}_{1}\cdot\hat{r}_{2})
\label{eqn:mp_N}
\end{equation}
and
\begin{equation}
RRR=\sum_{l'}\mathcal{R}_{l'}(r_{1},r_{2})P_{l'}(\hat{r}_{1}\cdot\hat{r}_{2})
\label{eqn:mp_R}
\end{equation}
then 
\begin{equation}
\frac{\mathcal{N}_{k}}{\mathcal{R}_{0}}=\zeta_{k}+\sum_{l}\zeta_{l}(2k+1)\sum_{l'>0}\left(\begin{array}{ccc}
l & l' & k\\
0 & 0 & 0
\end{array}\right)^{2}f_{l'}.
\label{eqn:edgecorrxn_fund}
\end{equation}
The $\zeta_l$ are the desired multipole coefficients of the 3PCF and the $f_{l'}\equiv \mathcal{R}_{l'}/\mathcal{R}_0$ are the ratios of higher multipoles of the randoms' triple count to its monopole. The Wigner 3j-symbol represents angular momentum coupling; it satisfies the Triangle Inequality $|l-l'|\leq k\leq|l+l'|$ and $l+l'+k$ must be even by parity.  

We see that a vector of measured $\mathcal{N}_k/\mathcal{R}_0$ is simply a linear transformation of the vector of $\zeta_k$ we desire, and given the matrix $\bf{I}+\bf{M}$ where $\bf{I}$ is the identity matrix and $\bf{M}$ has elements
\begin{equation}
M_{kl}=(2k+1)\sum_{l'>0}\left(\begin{array}{ccc}
l & l' & k\\
0 & 0 & 0
\end{array}\right)^{2}f_{l'},
\label{eqn:Mkl}
\end{equation}
the edge-correction equation becomes
\begin{equation}
\vec{\mathcal{N}}/\mathcal{R}_{0}=(\boldsymbol{I}+\boldsymbol{M})\vec{\zeta}\equiv\boldsymbol{A}\vec{\zeta},
\label{eqn:matrix_edgecorrxn}
\end{equation}
where $\vec{\mathcal{N}}=(\mathcal{N}_{0},\mathcal{N}_{1},\cdots,\mathcal{N}_{l_{\rm max}})$, and analogously for $\vec{\zeta}$, with $l_{\rm max}$ the maximum multipole to which we measure. This edge-correction equation can then be solved for $\vec{\zeta}$ by matrix inversion. In practice the $f_{l'}$ fall rapidly with $l'$, meaning we can measure them to $l'=10$ and compute edge-corrected $\zeta_l$ up to $l=9$, as discussed in SE15b.

It is necessary to use a large number of random points so that the
Poisson error from them is much smaller than the Poisson error from
the galaxy data.  We do this by computing $NNN$ with 1.5 times more
randoms than data, weighting the randoms so that the sum of their
weights is equal to that of the data. The calculation of $NNN$ is repeated with 32 different sets of randoms, 
and the results then averaged to get our final $NNN$; this is equivalent
to using $1.5\times 32 = 48$-fold more randoms than data when computing the data-random
pairs.  As described in SE15b, this process optimizes the minimization
of the Poisson error for a fixed amount of computational time, by
avoiding an expensive excess of random-random pairs.  When computing
the $NNN$ of mock catalogs, we use only 10 sets of randoms.  This incurs
slightly more Poisson error in the resulting covariance matrix, but
is 3-fold faster computationally.  We estimate $RRR$ and the resulting
$\mathcal{R}_0$ and $f_{l'}$ factors using simply one set of randoms,
1.5 times more numerous than the data.  This is sufficient for
accurate answers as these terms do not involve the difference of
two nearly canceling ingredients as $N$ does.

\section{Compression}
\label{sec:compression}
Using an idea first advanced in SE15a, we compress the radial coefficients $\zeta_l(r_1,r_2)$ by integrating over $r_2$ in a region fixed by $r_1$.  This has two motivations. 

\begin{figure}
\centering
\includegraphics[scale=0.42]{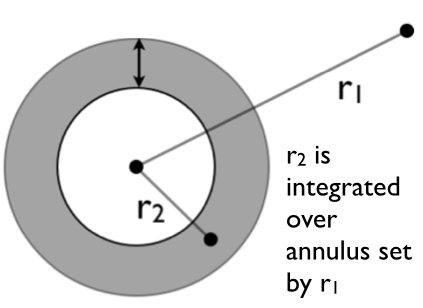}
\caption{A visualization of the compression scheme; dots represent galaxies, and $r_1$ and $r_2$ are the triangle sides. The grey annulus contains allowed $r_2$ values given a particular $r_1$. In the present work, the inner radius of the annulus is fixed to be $30\Mpch$ and the outer edge $r_1-30\Mpch$.}
\label{fig:comp_vis}
\end{figure}

First, the integration region for $r_2$ can be set so as to avoid the squeezed limit where $r_1=r_2$ and the third triangle side, $r_3$, can become zero.  In this squeezed limit, as $\theta_{12}\to 0$ the two galaxies separated by $r_3$ begin to interact with each other strongly via gravity and so perturbation theory will likely be invalid. Further, the 3PCF becomes large when one triangle side is small, so these squeezed, third-side-zero triangles actually dominate the isosceles $r_1 =r_2$ signal. By choosing an integration region in $r_2$ that constrains $r_2$ to be greater than some minimum and also well different from $r_1$, we can avoid this squeezed limit.  The  compression scheme is shown in Figure \ref{fig:comp_vis}.

Second, compression lowers the dimension of the problem, reducing the dimension of the covariance matrix, easing parameter fitting, and permitting simpler visualization of the data, as a 1-D line plot versus $r_1$ rather than a 2-D color map of $\zeta_l(r_1,r_2)$ amplitudes versus $r_1$ and $r_2$. This framework enables more robust interpretation and analysis of the 3PCF.

However, we believe that the
compression is not the optimal choice for extraction of BAO
information.  As SE15a shows, the full 3PCF in the multipole basis has structure where $r_3=r_1 + r_2$ equals the BAO scale of $100\Mpch$. This structure gets averaged out in this compression. In the present work we present the compressed approach as a first step, in that
it can validate the covariance matrix and the basic broadband
agreement with perturbation theory.  Even despite the non-optimality of the compression for this purpose, we find a BAO signal, and we will extend our fitting to the full two triangle side and multipole space soon.

We now present the explicit form of our compression scheme. The radial binning means the original compression scheme of SE15a, which took $r_1$ and $r_2$ as continuous variables, required modifications, described in SE15b.  The compression used in that work and here is 
\begin{align}
\zeta^{\rm c}_l(r_1) = \frac{\sum _{r_2\in S(r_1)} \zeta_l(r_1,r_2) \Delta V(r_2)}{\sum _{r_2\in S(r_1)} \Delta V(r_2)}
\end{align}
where superscript $c$ denotes compressed, $\Delta V(r_2)$ is the volume of bin $r_2$, and $S(r_1)$ is the set of all bins in $r_2$ where $r_2$ is greater than $3\Delta r$ and less than $r_1 - 3\Delta r$, with $\Delta r$ the bin width. For our bin width of $\Delta r = 10\Mpch$, this prescription guarantees that $r_2 >30\Mpch$ and $|r_2 - r_1|>30\Mpch$, meaning by the Triangle Inequality that $r_3>30\Mpch$. As an example, for $r_1 = 100 \Mpch$, the $r_2$ range integrated over is $30\Mpch - 70 \Mpch$, or four $r_2$ bins given our $\Delta r$. This compression scheme means that for $r_1<70\Mpch$, there are no allowed $r_2$ bins, so it places a minimum on $r_1$ as well.  

Overall, the compression guarantees that all the galaxies in the triangle are well-separated so PT should describe them well.  The framework of compression and its sensitivity to linear, non-linear, and baryon-dark matter relative velocity bias are further discussed in SE15a. SE15b shows how this compression scheme translates to the covariance matrix, which we will also briefly summarize when we turn to the covariance matrix in \S\ref{sec:covar}.

\section{Model}
\label{sec:model}
\subsection{Standard perturbation theory}
\label{subsec:pt_model}
We use standard perturbation theory (SPT) to compute the 3PCF at leading (fourth) order in the linear density field $\delta$. Summarizing the calculation performed in SE15a, we begin with the real-space local (Fry \& Gazta\~naga 1993) Eulerian galaxy bias model
\begin{align}
\delta_{{\rm g}}\left(\vec{r}\right)=b_{1}\delta_{{\rm m}}\left(\vec{r}\right)+b_{2}\left[\delta_{{\rm m}}^{2}\left(\vec{r}\right)-\left<\delta_{{\rm m}}^{2}\right>\right]
\label{eqn:gal_bias_model}
\end{align}
where $\delta_{\rm g}$ is the galaxy overdensity at $\vec{r}$, $b_1$ is the linear bias, and $b_2$ is the non-linear bias. We do not include a bias term coupling to the tidal tensor here. $\delta_{\rm m}$ is the matter density, given in terms of the linear density field as
\begin{align}
\delta_{\rm m}(\vr) = \delta(\vr) +\delta^{(2)}(\vr)
\end{align}
with $\delta^{(2)}$ the usual second-order density field, most often given in Fourier space as
\begin{align}
&\tilde{\delta}^{(2)}\left(\vk\right) =\nonumber\\
& \int d^3 \vk_1d^3\vk_2\; \delta_{\rm D}^{[3]}(\vk_1 +\vk_2-\vk)\tilde{F}^{(2)}(\vk_1,\vk_2)\tilde{\delta}(\vk_1)\tilde{\delta}(\vk_2)
\end{align}
with $\delta_{\rm D}^{[3]}$ a 3-D Dirac delta function and $\tilde{F}^{(2)}$ the second-order kernel
\begin{align}
&\tilde{F}^{(2)}(\vk_1,\vk_2) \nonumber\\
&= \frac{17}{21}P_0(\hk_1\cdot\hk_2) + \frac{1}{2}\left(\frac{k_1}{k_2} +\frac{k_2}{k_1}\right)P_1(\hk_1\cdot\hk_2) +\frac{4}{21}P_2(\hk_1\cdot\hk_2)
\label{eqn:f2_kernel}
\end{align}
(Bernardeau et al. 2002 equation (45)).

We calculate the 3PCF to fourth order using the bias model (\ref{eqn:gal_bias_model}) and focus on the pre-cyclic terms. This means that we choose one vertex of the triangle
of galaxies at which to define $\theta_{12}$ and the two sides enclosing
it, $r_{1}$ and $r_{2}$. Note that in the product of three copies
of $\delta_{g}$ needed to form the 3PCF, each of the three galaxies
can contribute a $\delta^{2}$ and a $\delta^{(2)}$.
The pre-cyclic term is written by choosing one galaxy to contribute
each of these (it may be the same one). We have chosen the third galaxy to
contribute these more complicated terms. Since we
can then take this galaxy to be at the origin, this approach simplifies
the calculation. However, to connect with observations, where there
is no ``preferred'' vertex (galaxy), we eventually must sum cyclically,
giving each galaxy in the survey the chance to contribute $\delta^{2}$
and $\delta^{(2)}$.

Denoting ``pre-cyclic'' with a subscript $pc$, we find
\begin{equation}
\zeta_{{\rm pc}}\left(r_1,r_2;\hr_1\cdot\hr_2\right)=\sum_{l=0}^{2}\zeta_{{\rm pc}l}\left(r_{1},r_{2}\right)P_{l}\left(\hr_1\cdot\hr_2\right),
\end{equation}
with the coefficients as
\begin{equation}
\zeta_{{\rm pc}0}\left(r_{1},r_{2}\right)=\left[2b_{1}^{2}b_{2}+\frac{34}{21}b_{1}^{3}\right]\xi_{1}\xi_{2},
\label{eqn:pc_ell0}
\end{equation}
\begin{align}
\zeta_{{\rm pc}1}\left(r_{1},r_{2}\right)=-b_{1}^{3}\left[\xi_{1}^{\left[1-\right]}\xi_{2}^{\left[1+\right]}+\xi_{2}^{\left[1-\right]}\xi_{1}^{\left[1+\right]}\right]
\end{align}
and
\begin{equation}
\zeta_{{\rm pc}2}\left(r_{1},r_{2}\right)=\frac{8}{21}b_{1}^{3}\xi_{1}^{\left[2\right]}\xi_{2}^{\left[2\right]}.
\end{equation}
Subscripts indicate sides $r_1$ or $r_2$ as a function's argument, and $\xi$ is the linear theory matter correlation function;
$\xi^{\left[2\right]}$ and $\xi^{\left[1\pm\right]}$ are defined as:
\begin{equation}
\xi^{[1\pm]}\left(r\right)=\int_{0}^{\infty}\frac{dk}{2\pi^{2}}k^{2}j_{1}\left(kr\right)P\left(k\right)k^{\pm1}
\label{eqn:xi_1}
\end{equation}
and
\begin{equation}
\xi^{[2]}\left(r\right)=\int_{0}^{\infty}\frac{dk}{2\pi^{2}}k^{2}j_{2}\left(kr\right)P\left(k\right),
\label{eqn:xi_2}
\end{equation}
where $P(k)$ is the matter power spectrum. 

We will consider two different matter power spectra: the physical power spectrum, denoted $P_{\rm phys}$, and a ``no-wiggle'' power spectrum without BAO, denoted $P_{\rm nw}$. This no-wiggle power spectrum is computed using Eisenstein \& Hu (1998)'s fitting formula for the no-wiggle transfer function, while the physical power spectrum is computed, following Eisenstein, Seo \& White (2007) and Anderson et al. (2012), as 
\begin{align}
P_{\rm phys} (k) = \left[  P(k) - P_{\rm nw}(k)\right]\exp[-k^2\Sigma_{\rm nl}^2/2] + P_{\rm nw}(k),
\label{eqn:p_template}
\end{align}
where $\Sigma_{\rm nl}=8\Mpch$ is a non-linear smoothing scale to represent non-linear structure formation and RSD.

We pause to note an important feature of equation (\ref{eqn:xi_1}) regarding the BAO. The additional factor of $k$ in $\xi^{[1+]}$ amplifies the sharp, localized real-space BAO bump encoded in the power spectrum (SE15a).  This factor of $k$ stems from the $\ell=1$ term in the $\tilde{F}^{(2)}$ kernel (\ref{eqn:f2_kernel}), which in turn is generated primarily ($\sim70\%$) by density gradients parallel to the velocity (SE15b; Bernardeau et al. 2002). The density Green's function (response to an initial Dirac-delta function overdensity at high redshift) has a sharp BAO bump (Slepian \& Eisenstein 2015d), and so its divergence goes from positive to negative with a zero crossing at the BAO scale. Multiplying by the velocity amplifies structure at and beyond the BAO scale relative to that on much smaller scales because the velocity is larger at and beyond the BAO scale (SE15a). The remaining $30\%$ of the $\ell=1$ term in $\tilde{F}^{(2)}$, which also enters $\xi^{[1+]}$, stems from gradients of the velocity divergence parallel to the velocity, which also contain BAO information. Mathematically, the weighting by $k$ in equations (\ref{eqn:f2_kernel}) and (\ref{eqn:xi_1}) occurs because the gradient operator's Fourier Transform scales as $k$. As \S\ref{sec:results} will show, these points lead to an order-unity BAO feature in the weighted, compressed 3PCF's dipole $(\ell = 1)$.

After cyclically summing,
we re-project onto the basis of Legendre polynomials to find the full radial coefficients in our multipole expansion (\ref{eqn:3pcf_multi}) of the 3PCF as
\begin{align}
\zeta_{l}\left(r_{1},r_{2}\right)&=\frac{2l+1}{2}\int_{-1}^{1}d\mu_{12}\bigg[\zeta_{{\rm pc}}\left(r_{1},r_{2},\mu_{12}\right)\\
&+\zeta_{{\rm pc}}\left(r_{2},r_{3},\mu_{23}\right)+\zeta_{{\rm pc}}\left(r_{3},r_{1},\mu_{31}\right)\bigg]P_l(\mu_{12}),\nonumber
\end{align}
with $\mu_{12}\equiv\cos\theta_{12}$ and analogously for $\mu_{23}$ and $\mu_{31}$. Note that $r_{3},\;\mu_{23}$ and $\mu_{31}$ are all functions of
$r_{1,}\; r_{2},$ and $\mu_{12}$, easily found using the law of
cosines. The factor of $\left(2l+1\right)/2$ is necessary because
$\int_{-1}^{1}P_{m}\left(\mu\right)P_{n}\left(\mu\right)d\mu=2/\left(2n+1\right)\delta^{\rm K}_{mn}$;
that is, the Legendre polynomials are an orthogonal but not orthonormal
basis. Where in the pre-cyclic terms, we only found terms up to $l=2$, cyclically summing introduces higher orders. These  coefficients are shown versus $r_1$ and $r_2$ in SE15a Figure 9 split out by bias coefficient; since we set $b_v\equiv 0$, only the first two columns of this three-column figure are relevant to the present analysis. Also note that this Figure is in ${\rm Mpc}$ whereas the present work uses$\Mpch$.

The power spectrum used in these predictions is computed from the matter transfer function  $T_{\rm m}$ given by CAMB (Lewis 2000). We have $P = Ak^{n_s}T_{\rm m}^2$, with $A$ a normalization set by matching $\sigma_8$ and $n_s$ the scalar spectral tilt. The transfer function is on a grid equally spaced in $\log k$ with $5000$ divisions per decade from $k=3.46\times 10^{-6}$ to $1.87\;h/{\rm Mpc}$.  To avoid ringing in the transforms of equations (\ref{eqn:xi_1}) and (\ref{eqn:xi_2}) we multiply by a Gaussian smoothing in Fourier space $\exp[-k^2]$, which suppresses structure below $1\Mpch$. The predictions are not sensitive to the exact value of the smoothing scale.

We use a geometrically flat $\Lambda{\rm CDM}$ cosmology with parameters matching those used for the \textsc{MultiDark-Patchy} mock catalogs (Kitaura et al. 2015): $\omegab = 0.048,\; \omegam = 0.307115, \; h\equiv H_0/(100\;{\rm km/s/Mpc})=0.6777, \; n_s = 0.9611, \;\sigma_8(z=0) = 0.8288,\; T_{\rm CMB} = 2.7255\;{\rm K}$. These parameters do not differ substantially from the Planck values (Planck Paper XIII, 2015). We rescale $\sigma_8$ by the ratio of the linear growth factor at the survey redshift to the linear growth factor at redshift zero. We take the survey redshift to be $z_{\rm survey}=0.565$ by simply averaging $0.43$ and $0.7$; since the redshift distribution of objects is roughly symmetric about the middle of the redshift interval this is a good approximation.  

Broadly, it is worth emphasizing that the PT model we use is fourth order in the linear density field: this is the lowest order at which the 3PCF is non-zero.  It does not incorporate any modeling of spherical collapse and virialization. However, on the large scales $(\gtrsim 30\Mpch)$ we present here, it is likely that such effects are not important, especially because our compression scheme (\S\ref{sec:compression}) prevents any triangle side from becoming smaller than $30\Mpch$.  

%%%%
\subsection{Redshift-space distortions}
\label{subsec:rsd}
Our model does not include any treatment of the effects of Redshift-Space Distortions (RSD). In principle, RSD can generate additional angular and side-length-dependent structure in the 3PCF. In practice, however, we believe that in our compressed basis the dominant effect of RSD is simply to rescale the multipole moments by some overall factor that depends only mildly on side length and multipole (SE15b). Consequently the no-RSD PT model presented above can fit the data well, but the bias parameters appearing in it are actually redshift-space quantities rather than the usual real-space biases.  

In SE15b, we measured the real-space and redshift-space compressed 3PCF's multipole moments on scales from $45-90\Mpch$ for the LasDamas SDSS DR7 mock catalogs.  SE15b Figure 12 shows that RSD rescale the  3PCF by a factor of $\sim 1.5-1.8$ roughly independent of side length and multipole.  Previous analytic work on the bispectrum by Scoccimarro, Couchman \& Frieman (1999; hereafter SCF99) has shown that the rescaling is mostly set by $\beta = f/b_1$, where $f=d\ln D/d\ln a\approx \Omega_{\rm m}^{0.55}$ is the logarithmic derivative of the linear growth rate $D$ with respect to scale factor $a$ and $b_1$ is the real-space linear bias. For the LasDamas mocks we found $\beta = 0.33$; using the real-space linear bias $b_1=2.086$ from the CMASS power spectrum (Gil-Mar\'in et al. 2015), $\beta =  0.37$ for the present work. Given the similarity of the CMASS and LasDamas values of $\beta$, we expect that the impact of RSD on the CMASS compressed 3PCF is comparable to that shown in SE15b Figure 12. 

Importantly, since at fixed  $\gamma \equiv 2b_2/b_1$ the 3PCF is proportional to $b_1^3$, any overall rescaling of the 3PCF due to RSD will only enter the measured redshift-space $b_1$ as a cube root.\footnote{The term in $b_1^2b_2$ of equation (\ref{eqn:pc_ell0}) can be rewritten as $b_1^3\gamma$. $\gamma$ is defined following SCF99, but their real-space $b_2$ is defined such that it is twice the real-space $b_2$ of the present work.} Thus rescaling factors in the range $1.5-2$ will only change the implied real-space $b_1$ by $14-26\%$. 

In detail, the rescaling due to RSD does also depend explicitly on $b_1$. The real-space linear bias $b_1 = 1.90$ for LasDamas, comparable to the value $b_1=2.086$ Gil-Mar\'in et al. (2015) find for CMASS, so the $b_1$-dependence is unlikely to make the effect of RSD different between the two. The rescaling also depends on $\gamma$; for LasDamas we found $\gamma = 0.99$, while using the Gil-Mar\'in et al. (2015) value for $b_2$ from the power spectrum we have $\gamma = 0.43$ for CMASS. While these values differ, the dependence on $\gamma$ is weak, $\sim 3\%$ in $\ell=0$ and $\ll 1\%$ for $\ell=1$ and $2$ for a factor of $2$ change in $b_2$. Thus the empirical result that RSD rescale the LasDamas compressed 3PCF roughly independently of side length and multipole is likely to hold for the CMASS compressed 3PCF as well.

We now offer physical intuition as to why this rescaling is roughly a constant.  One can show analytically that the rescaling is independent of side length at $\oO(\beta)$ and nearly so at $\oO(\beta^2$). Over a realistic range of $b_1$ and $b_2$ the variation with multipole is also modest. Given that $\beta^3\sim 3\%$, higher-order corrections might be expected to be unimportant.  

In more detail, at $\oO(\beta)$, simplifying SCF99 we find the following rescaling factors:
\begin{align}
\frac{\zeta_{0,{\rm s}}}{\zeta_0} & = 1+ \frac{94+42(b_1+\gamma)}{102+63\gamma}\beta,\nonumber\\
\frac{\zeta_{1,{\rm s}}}{\zeta_1} & = 1 + \left(1+\frac{b_1}{3}\right)\beta,\nonumber\\
\frac{\zeta_{2,{\rm s}}}{\zeta_2} & = 1 + \frac{4}{3}\beta.
\label{eqn:lo_beta_rsd}
\end{align}
Subscript $s$ denotes redshift-space.  At $\oO(\beta^2)$, simple analytic formulae are not available but numerically taking the multipole moments of the SCF99 expression for the bispectrum (averaged over all orientations with respect to the line of sight) shows that dependence on the magnitudes of $k_1$ and $k_2$ is negligible. We have used these $\oO(\beta^2)$ results to estimate the rescaling expected for the LasDamas compressed 3PCF. For $\ell=0,1$, and $2$, the empirical rescalings from SE15b are $1.50,\;1.71$ and $1.72$. Our $\oO(\beta^2)$ results using SCF99 are $1.52,\;1.70$ and $1.80$, which differ from the SE15b values by respectively $1.2\%,\;-0.59\%$, and $4.7\%$. This agreement is encouraging and suggests that our approach could be used to relate redshift-space and real-space measurements to reasonable accuracy. For the CMASS sample, we estimate the rescalings are $1.74,\;1.87$, and $1.99$ for $\ell = 0,\;1$ and $2$. While we expect these factors are accurate at the $\sim 5\%$ level, we defer detailed modeling of RSD to future work. Here we do not use these estimated rescalings to recover real-space biases from our redshift-space fits, but simply report the redshift-space best fit biases for the model equation (\ref{eqn:gal_bias_model}). 

We now summarize the approaches to RSD of other recent 3PCF and bispectrum works. For the bispectrum we focus on Gil-Mar\'in et al. (2015), the most recent work on the bispectrum and the one using the largest sample to date. This work uses the full bispectrum model of SCF99 with an additional overall damping factor to describe fingers-of-God (Jackson 1972) from velocity dispersions inside virialized structures. The SPT kernels of SCF99 are also modified to ``effective'' kernels, though this is to better describe the smaller-scale, weakly non-linear regime rather than to address RSD. Gil-Mar\'in et al. (2015) also add a non-local bias term, which effectively rescales the $P_2$ and $P_0$ terms in the $\tilde{F}^{(2)}$ kernel.

For the 3PCF, we focus on Gazta\~naga et al. (2009), which measures the 3PCF on BAO scales, and on McBride et al. (2011a,b), the largest sample used for the 3PCF previous to the present work. Gazta\~naga et al. (2009) measure the reduced 3PCF, which is the 3PCF divided by the hierarchical ansatz that $\zeta(r_1,r_2,r_3)\sim \xi(r_1)\xi(r_2)+{\rm cyc.}$, with $\xi$ the linear correlation function (Groth \& Peebles 1977).  They note that on large scales RSD cancel out of the reduced 3PCF, as shown in their Figure 2, so they do not incorporate additional modeling for RSD.  The bias model of McBride et al. (2011a,b) has the reduced 3PCF of galaxies as a constant rescaling of the dark matter-only 3PCF. To remove the effects of RSD from their bias measurement, McBride et al. (2011a,b) apply a shift to the dark matter particles in the Hubble Volume simulations used for the dark matter-only 3PCF.  As McBride et al. (2011a) Figure 8 shows, this is accurate save on the smallest scales they measure ($\lesssim  9\Mpch$).  

Finally, the \textsc{MultiDark-Patchy} mocks used for parts of the present analysis have been tuned to match small-scale quasi-virialized motions and the concomitant fingers-of-God, but for triangle opening angles $\theta=0$ and $\pi$ do not reproduce the observed 3PCF well (Kitaura et al. 2015; Figure 6).  However this mismatch is for a triangle with $r_1,\;r_2=10,\;20\Mpch$, and on the larger scales in the present analysis we expect that fingers-of-God will not be an important effect. In particular, the compression scheme guarantees that all triangle sides are greater than $30\Mpch$, and further averages two triangle sides ($r_2$ and $r_3$) over rather large bins, additionally suppressing any finger-of-God effect.

%%%%

\subsection{Integral Constraint Model}
\label{subsec:int_const}
We consider whether the randoms used for edge-correction may have been incorrectly normalized and hence fail to satisfy the integral constraint, that the mean number density of randoms matches the true number density of galaxies.  Such a misnormalization can occur because we only know the measured number density of galaxies, which may be different from the true number density were one to perform the survey over an infinite volume.  

We incorporate a possible failure of the integral constraint into the PT model fit to the data by adding a small fractional shift $c$ to the randoms, taking our 3PCF estimator from 
\begin{align}
\zeta = \frac{(D-R)^3}{RRR}\to\frac{[D-R(1+c)]^3}{RRR(1+c)^3}.
\end{align}
Expanding this out and simplifying we find
\begin{align}
\zeta^{\rm M} = \frac{\zeta^{\rm T} -c\xi_{{\rm cyc}} - c^3}{(1+c)^3},
\end{align}
where $\zeta^{\rm M}$ is the measured 3PCF and $\zeta^{\rm T}$ is the true 3PCF.  $\xi_{{\rm cyc}}$ is the cyclic sum of the linear theory 2PCF around the three triangle sides, 
\begin{align}
\xi_{{\rm cyc}}(r_1,r_2,r_3)\equiv \xi(r_1) +  \xi(r_2) +  \xi(r_3).
\label{eqn:xi_cyc}
\end{align}

The 2PCF fits have often included similar nuisance polynomials, including a constant in $\xi$ (e.g. Anderson et al. 2012). Including this constant in the BAO fits helps make them robust to observational systematics (Osumi et al. 2015; Ross et al. 2012; Ross et al. 2015). This term also is a good approximation of anomalous systematic power at small wavenumber.

\section{Covariance Matrix}
\label{sec:covar}
\subsection{A Hybrid Approach}
The covariance matrix {\bf C} measures the independence of different multipole and radial bin combinations from each other. In the limit of fully independent measurements {\bf C} is diagonal.  The $\chi^2$ for a parameter fit is calculated via $\vec{v} {\bf C}^{-1}\vec{v}^T$, where $\vec{v}$ is a vector of the data points minus the model being fit.  Minimizing the $\chi^2$ according to this prescription is guaranteed to give the optimal parameter fit if the likelihood is Gaussian.

However, this approach requires that the covariance matrix be invertible. Consequently if the covariance matrix is determined empirically by measuring the 3PCF of mock catalogs, a large number of mocks for each matrix element is desirable (Percival et al. 2014). Given the large dimension of the covariance matrix this is computationally costly, and so we have adopted an alternative approach. We compute the covariance matrix analytically and treat the effective survey volume and number density as free parameters to be fit by matching the empirical covariance derived from mock catalogs. This empirical covariance matrix, while likely too noisy to invert, is sufficient to constrain the volume and shot noise and reveal whether our analytic calculation describes the survey well.

Assuming that the density field is purely Gaussian and that the survey is boundary-free, the covariance of our multipole decomposition of the 3PCF can be analytically computed as (SE15b)
\begin{align}
&C_{{\rm GRF}, ll'}(r_{1},r_{2};r_{1}',r_{2}')=\frac{4\pi}{V}(2l+1)(2l'+1)(-1)^{l+l'}\nonumber\\
&\times\int r^{2}dr\sum_{l_{2}}(2l_{2}+1)\left(\begin{array}{ccc}
l & l' & l_{2}\\
0 & 0 & 0
\end{array}\right)^2\nonumber\\
&\times\bigg\{(-1)^{l_2}\xi(r)\bigg[f_{l_{2}ll'}(r;r_{1},r_{1}')f_{l_{2}ll'}(r;r_{2},r_{2}')\nonumber\\
&+f_{l_{2}ll'}(r;r_{2},r_{1}')f_{l_{2}ll'}(r;r_{1},r_{2}')\bigg]+(-1)^{(l+l'+l_{2})/2}\nonumber\\
&\times\bigg[f_{ll}(r;r_{1})f_{l'l'}(r;r_{1}')f_{l_{2}ll'}(r;r_{2},r_{2}')\nonumber\\
&+f_{ll}(r;r_{1})f_{l'l'}(r;r_{2}')f_{l_{2}ll'}(r;r_{2},r_{1}')\nonumber\\
&+f_{ll}(r;r_{2})f_{l'l'}(r;r_{1}')f_{l_{2}ll'}(r;r_{1},r_{2}')\nonumber\\
&+f_{ll}(r;r_{2})f_{l'l'}(r;r_{2}')f_{l_{2}ll'}(r;r_{1},r_{1}')\bigg]\bigg\},
\label{eqn:C_GRF}
\end{align}
with
\begin{equation}
f_{ll}(r;r_{1})=\int\frac{k^{2}dk}{2\pi^{2}}\left[P(k)+\frac{1}{n}\right]j_{l}(kr_{1})j_{l}(kr)
\label{eqn:f2_tensor}
\end{equation}
and
\begin{equation}
f_{l_{2}ll'}(r;r_{1},r_{1}')=\int\frac{k^{2}dk}{2\pi^{2}} \left[P(k)+\frac{1}{n}\right] j_{l}(kr_{1})j_{l'}(kr_{1}')j_{l_{2}}(kr),
\label{eqn:f3_tensor}
\end{equation}
where $V$ is the effective survey volume, and  $n$ is the effective number density of the survey. The $1/n$ term added to the power spectrum above characterizes the shot noise contribution to the covariance matrix. The covariance as above can be easily binned radially and compressed by integrating out $r_2$ and $r_2'$, as detailed in SE15b. 

Note that linear perturbation theory is insufficient to obtain the 3PCF for a purely Gaussian Random Field, because as an odd moment the 3PCF vanishes. However, linear theory does give the covariance because it is a 6-point function and hence an even moment. An analogy here is to a distribution with zero mean (the expected signal, the 3PCF), but non-zero width (the covariance).
 
To fit for volume and shot noise, we compute the empirical covariance matrix with elements
\begin{align}
&C_{{\rm mock},ll'}(r_1,r_2; r_1', r_2')= \nonumber\\
& \frac{1}{N_{\rm mock}-1}\sum_{i=1}^{N_{\rm mock}} \left[\zeta_{i,l}(r_1,r_2)-\left< \zeta_{l}(r_1,r_2)\right>\right]\nonumber\\ &\times\left[\zeta_{i,l'}(r_1',r_2')-\left< \zeta_{l'}(r_1',r_2')\right>\right]
\end{align}
where $N_{\rm mock}$ is the number of mock catalogs, $i$ denotes the mock catalog number, and angle brackets represent the mean over the mocks.

\begin{figure}
\centering
\includegraphics[scale=1]{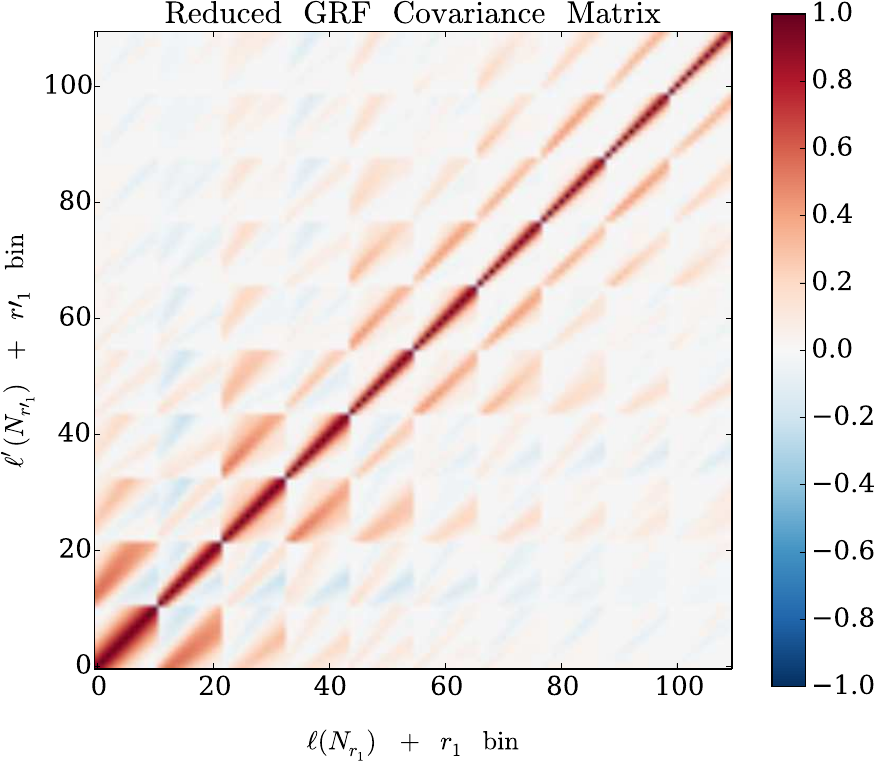}
\caption{The reduced covariance matrix from our analytic calculation, ${\bf C}_{{\rm GRF},ll'}(r_1,r_1')/\sqrt{{\bf C}_{{\rm GRF},ll}(r_1,r_1){\bf C}_{{\rm GRF},l'l'}(r_1',r_1')}$. We have mapped the 4-D compressed covariance  tensor, which depends on $r_1, r_1', l$ and $l'$, to 2-D as indicated by the axis labels. Each small $11\times11$ tile corresponds to fixed $l$ and $l'$; $N_{r_1}=N_{r'_1}=11$ is the number of bins in side-length. The tile beginning at the origin represents the covariance of all radial bins at $l=l'=0$. The two tiles neighboring the $l=l'=0$ tile represent $l=1,\;l'=0$ and $l'=1,\;l=0$. The strong diagonal band within many of the tiles represents the covariance of radial bins where $r_1 =r_1'$.}
\label{fig:reduced_covar}
\end{figure}

The metric used for fitting the volume and shot noise of our analytic calculation to the empirical covariance is given by Xu et al. (2012) equation (58).  For a fixed shot noise $n$, the likelihood-maximizing volume $V_{\rm best}$ can be found analytically as 
\begin{align}
V_{\rm best} = \frac{d}{\left< \chi_1^2\right>}
\end{align}
where $d$ is the dimension of the covariance matrix. $\left< \chi_1^2\right>$ is the mean $\chi^2$ of all the mock covariances with respect to the analytic covariance matrix with the given shot noise and with the volume set to unity.  Explicitly,
\begin{align}
\left< \chi_1^2\right> = \sum_{i=0}^{N_{\rm mock}} \vx \;{\bf C}^{-1}_{1, \rm GRF}\;\vx^T
\end{align}
where ${\bf C}_{1, \rm GRF}$ is the analytic covariance matrix with $V=1$ and a given shot noise and $\vx$ is a vector of the differences between the mock multipoles $\zeta_l(r_1,r_2)$ and their mean taken over all mock catalogs. We find a best-fit volume of $1.92\;{\rm Gpc}^3/h^3$ and a best-fit number density of $n=1.80\times10^{-4}\;h^3/{\rm Gpc}^3$. Figure \ref{fig:reduced_covar} shows the reduced covariance matrix used in our parameter fitting, i.e. that given by equation (\ref{eqn:C_GRF}) after compression and with the values of $V$ and $n$ quoted above. Away from the diagonal, the covariance is strongest for low multipoles, indicated by the higher amplitude in the lower left corner.

\subsection{Covariance Results and Tests}
To assess how well our analytic covariance calculation fits the mock covariance, we compare the eigenvalues of the two matrices. We also compare the ratio of the eigenvectors at each eigenvalue.  We find good agreement in both these tests. The agreement worsens somewhat as the eigenvalues become smaller, as we might expect because these will be more sensitive to the Poisson noise of estimating the covariance matrix using 299 mock catalogs rather than a larger number.  

\begin{figure}
\centering
\includegraphics[scale=1]{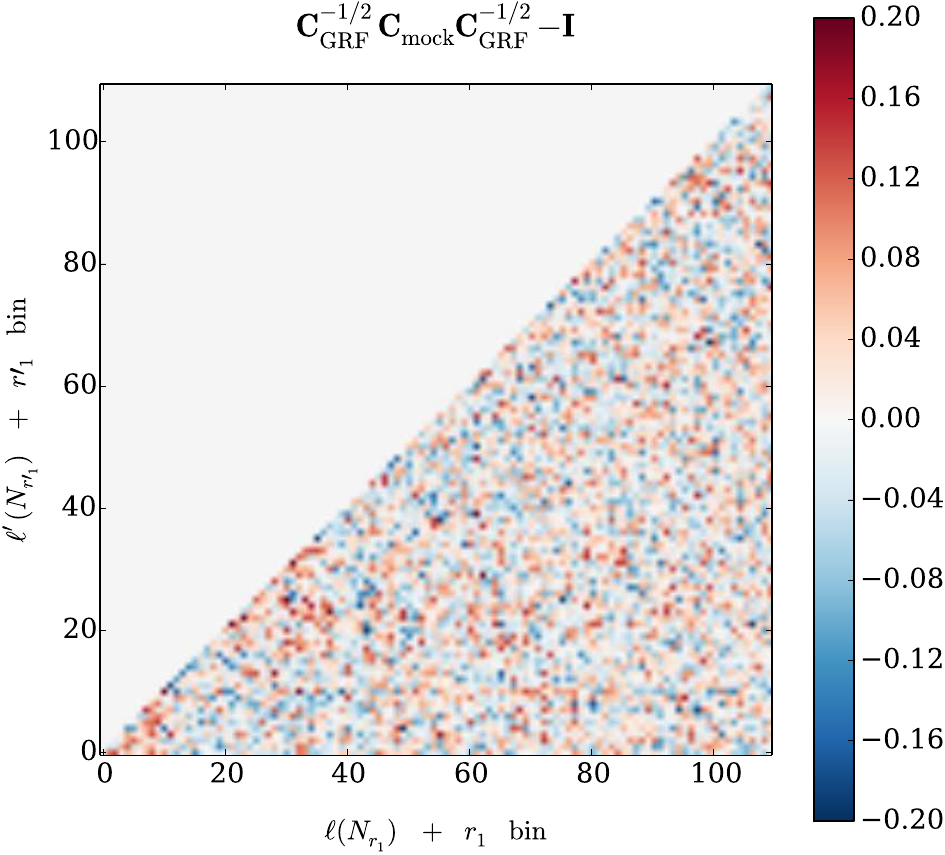}
\caption{A test of the analytic covariance $\bf{C}_{\rm GRF}$ against that estimated from the mocks, $\bf{C}_{\rm mocks}$. The mean of this plot is $0.6\%$ with root mean square $6\%$. The fact that the mean is close to zero shows that our analytic covariance matrix describes the true covariance well; $6\%$ is the rms expected with $110$ degrees of freedom and $299$ mocks, as discussed in the main text. We have blocked out the upper half because the matrix is symmetric by construction and this can create the misleading appearance by eye of more pattern than there is in reality. The mapping of the 4-D compressed covariance tensor to 2-D is the same as in Figure \ref{fig:reduced_covar}.}
\label{fig:covar_test}
\end{figure}

As a third test, we compute $\bf{C}_{\rm GRF}^{-1/2} \bf{C}_{\rm mock}\bf{C}_{\rm GRF}^{-1/2}-\bf{I}$, where $\bf{I}$ is the identity matrix. Note we apply a half-inverse on each side of the mock matrix purely to ensure our resulting test is symmetric. If the analytic and mock covariances are identical, this should be zero; given the noise of using a finite number of mocks in reality we expect the result to look like noise. This test is shown in Figure \ref{fig:covar_test}. The mean is $0.6\%$ and the root mean square is $6\%$. This scatter about the (roughly zero) mean is attributable to having used a finite number of mocks, as with $110$ degrees of freedom and $299$ mocks we expect rms of $110/\sqrt{299}\approx 6\%$.

Figure \ref{fig:covar_test} shows that our analytic calculation captures the true covariance as estimated from the mocks fairly well. We have windowed off the upper half of the test because the symmetry about the diagonal, present by construction, can misleadingly produce the appearance of more pattern by eye than is real.  Note that we have mapped the 4-D compressed covariance matrix ${\bf C}_{ll'}(r_1,r_1')$ into a 2-D array using row-major ordering along the multipoles; $l=0=l'$ is in the lower left corner and the radial bins vary faster than the multipoles.

While there are 2048 \textsc{MultiDark-Patchy} mocks extant, 299 mocks were sufficient for the present analysis. These mocks are only used to solve for two parameters (survey volume and shot noise), so the system is well-constrained algebraically.  Note that even all 2048 mocks would be insufficient if one sought to determine the full covariance matrix empirically, as it has $110^2/2=6,050$ independent elements, many more than one per mock.

\subsection{Details of Analytic Covariance Matrix Numerical Implementation}
Here we briefly detail our numerical implementation of the compressed version of equations (\ref{eqn:C_GRF})-(\ref{eqn:f3_tensor}).
As discussed in SE15b, the covariance can be efficiently compressed following the prescription of \S\ref{sec:compression} by compressing the $f$ tensors equations (\ref{eqn:f2_tensor}) and (\ref{eqn:f3_tensor}) first when their arguments include either $r_2$ or $r_2'$ or both.  However for some terms in the covariance the tensors only have $r_1$ and $r_1'$ as arguments, meaning we still need the full uncompressed tensors. We form these tensors on a linearly spaced grid in $k$ with $3000$ points from $10^{-4}\;h/{\rm Mpc}$ to $2\;h/{\rm Mpc}$. These tensors also involve a dummy variable $r$ which is integrated over at the end of the covariance calculation (see equation (\ref{eqn:C_GRF})); for this we use a linearly spaced grid with $4100$ points from $10^{-5}\Mpch$ to $1000\Mpch$.  A number of different endpoints for each grid and number of points for each grid were tested and the results found to be well-converged for the choices above. 

Binning adds an additional complication to the covariance calculation; as discussed in SE15b, one can analytically bin-average the spherical Bessel functions within the $f$ tensors (equations (\ref{eqn:f2_tensor}) and (\ref{eqn:f3_tensor}))  to incorporate this. In practice we found that simply numerically averaging them over the required bins was more accurate. To obtain the $f$ tensors, we use \textsc{Python's} internal implementation of the spherical Bessel functions to produce lookup tables; these need only be computed once even if one wishes to test a variety of cosmological parameter values in the 3PCF predictions.

Finally, for the covariance matrix, we wish to match as closely as possible the observed
redshift-space galaxy power spectrum.  We do so by rescaling $kP(k)/(2\pi^2)$ to be $90\; h^{-2}\;{\rm Mpc}^2$ at its maximum to match the observed CMASS power spectrum (Anderson et al. 2014).  This normalization incorporates redshift-space distortions and linear bias as they enter the power spectrum; empirically we find the required rescaling factor is $4.43$.
%Anderson et al. 2014 Figure 10, page 16.

\section{Results}
\label{sec:results}
\subsection{Mock and Data Results}
\label{subsec:md_results}
For both the CMASS data and the 299 \textsc{MultiDark-Patchy} mocks, we used the algorithm of SE15b to measure the 3PCF in 18 radial bins of $10\Mpch$ width each, covering triangle sides from $0$ to $180\Mpch$. Our compression scheme means that there are no allowed $r_2$ bins for $r_1<70\Mpch$ (\S\ref{sec:compression}), so after compression we have $11$ bins for $r_1$ going from $70\Mpch$ to $180\Mpch$. The run time was 2 hours per mock on
a 36-core node, about 20k core-hours for the full computation.  Computing the 3PCF of data and mocks was by far the dominant time cost of the calculation, with all other calculations taking negligible time. These include computing the covariance matrix and fitting for the best volume and shot noise, as well as fitting redshift-space linear and non-linear bias and the integral constraint parameter $c$ to the data and mock results.

\begin{figure}
\centering
\includegraphics[scale=1]{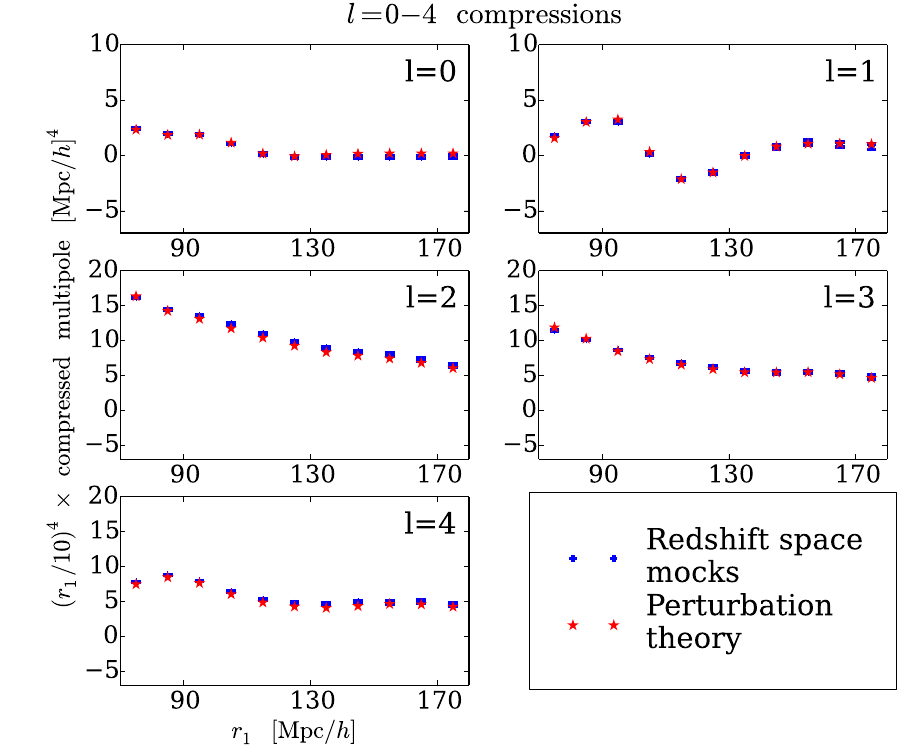}
\caption{Fit of PT predictions, computed using the physical power spectrum, to the compressed 3PCF's multipoles $\ell = 0-4$ for the mean of 299 \textsc{MultiDark-Patchy} mocks.  The BAO scale is $r_1=100\Mpch$; note the pronounced peak and trough around this scale in $\ell=1$ and the modest bumps in $\ell=0$ and $4$.  The tiny error bars on the mocks' mean are shown; they are the covariance's diagonal scaled down as appropriate for the mean of $299$ mocks, as further discussed in the main text.}
\label{fig:patchy_w_bao_lowell}
\end{figure}

\begin{figure}
\centering
\includegraphics[scale=1]{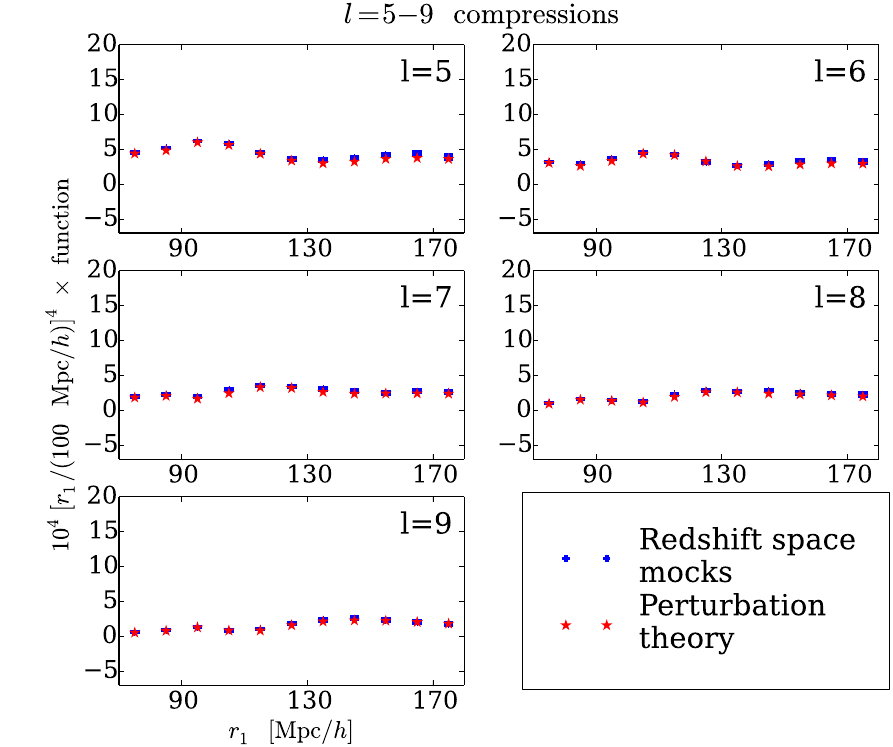}
\caption{Same as Figure \ref{fig:patchy_w_bao_lowell} but now for multipoles $\ell = 5-9$.  Notice the clear bump at the BAO scale of $r_1=100\Mpch$ in both $\ell=5$ and $6$. Note the similarity of these higher $\ell$ panels to each other, and indeed to $\ell = 4$ in Figure \ref{fig:patchy_w_bao_lowell}. This similarity indicates there is less information in the higher multipoles, for reasons further discussed in \S\ref{sec:basis} and S\ref{subsec:pt_model}. Essentially, prior to cyclic summing, the leading-order 3PCF has only $\ell=0,1,$ and $2$ terms, and so in our model higher multipoles stem from the purely geometric effect of cyclic summing. RSD can in principle introduce higher-multipole angular structure even at the pre-cyclic level, but  it seems this effect is not substantial.}
\label{fig:patchy_w_bao_hiell}
\end{figure}

For all fits, we use 10 multipoles, from $l=0$ to $9$, and 11 radial bins at each multipole. These fits for the mean of the $299$ \textsc{MultiDark-Patchy} mocks are displayed in Figures \ref{fig:patchy_w_bao_lowell} and \ref{fig:patchy_w_bao_hiell}. The error bars plotted are from the diagonal of the covariance matrix divided by $\sqrt{299}$ as is appropriate for a measurement of the mean of $299$ mocks. We have multiplied the compressed 3PCF multipoles by $(r_1/10)^4$ to remove the large-scale fall-off of the 3PCF, which from the hierarchical ansatz we expect to scale like the square of the 2PCF, $\zeta\sim \xi^2 \propto r^{-4}$ (Groth \& Peebles 1977). Notice the good match between the data and the PT. In particular, note the features at the BAO scale of $r_1=100\Mpch$: a significant peak and trough around the BAO scale in $\ell=1$, and bumps in $\ell=0$ and $4$.  As discussed in \S\ref{subsec:pt_model}, we expect the BAO to be an order-unity fractional change in the dipole $\ell=1$, and this is indeed borne out.

This fit shows that lowest-order PT in our compressed multipole basis appears to describe the 3PCF of these mocks fairly well. In particular, it supports our earlier claim that the primary effect of RSD in this compressed basis is a roughly multipole and side-length-independent rescaling of the 3PCF. 

Nonetheless, the fit's $\chi^2/{\rm d.o.f.} = 405.53/107=3.79 >1$, meaning that the PT is not a perfect match to these 299 mocks. We suspect this imperfect fit stems both from higher-order corrections to our leading-order PT model as well as possible scale or multipole-dependent effects of RSD. Nonetheless, this $\chi^2$ for the mocks suggests that the model should be more than sufficient to fit the data well. Given that $\chi^2\propto {\bf C}^{-1}\propto V$, we expect the model only begins to be insufficient for surveys with $V\sim 80{\rm x}$ that of SDSS DR12. Indeed, the PT model is an adequate fit to the data as we will soon show.

Table 1 gives our best-fit values of $b_1$, $b_2$, and $c$ for the mean of the 299 mocks. Again note these biases are redshift-space quantities. The error bars quoted in Table 1 are from integrating over the probability distribution given by $P\propto \exp[-\chi^2/2]$ to compute $\sigma_{b_1} = \sqrt{\left<b_1^2 \right>-\left<b_1\right>^2}$ from $\left< b_1\right>$ and $\left< b_1^2\right>$ and the same for $b_2$ and $c$.  They are the error bars we might expect if we did the measurement on a volume 299 times that of the SDSS DR12 CMASS sample. To verify our analysis procedure, we also examined the probability contours in $b_1-b_2$ space having marginalized over $c$, checking that the contours were consistent with our quoted error bars on $b_1$ and $b_2$. We do not show this plot here though we will display it for the CMASS data. 

Finally, Rodr\'iguez-Torres et al. (2015; see Figure 14) measure the linear bias for a set of 100 \textsc{MultiDark-Patchy} mocks using the 2PCF on scales up to $60\Mpch$. That work's model involves only linear bias and uses the Kaiser (1987) formula to remove the effects of RSD. They find some scale-dependence to the bias; the largest scale they measure is that most directly comparable to the present work. At this scale, the real-space $b_1 \simeq 1.8$, comparable to our inferred real-space bias of $1.90-2.00$ if the rescaling factors of \S\ref{subsec:rsd} are applied to the redshift-space mock bias $b_1=2.390$. The difference between our inferred real-space $b_1$ and Rodr\'iguez-Torres et al.'s measured value should not be over-interpreted for several reasons. First, the RSD factors used to convert our redshift-space bias are approximate; second, the bias model of Rodr\'iguez-Torres et al. (2015) includes only linear bias (ours includes non-linear bias as well), and third, they compute the bias by comparing the measured 2PCF to that from the dark matter halos, while we compare the measured 3PCF to the PT model of \S\ref{subsec:pt_model}.

\begin{figure}
\centering
\includegraphics[scale=1]{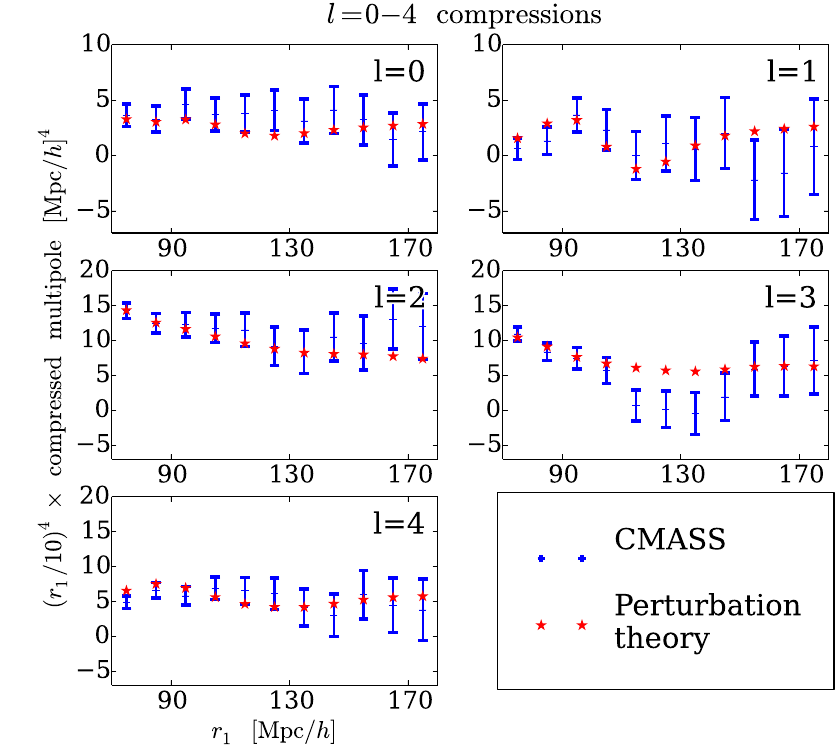}%updated. caption updated. 21 nov.
\caption{A fit of PT predictions, computed using the physical power spectrum, to the compressed 3PCF's multipoles $\ell = 0-4$ for the CMASS sample. Notice the peaks in $l=0, 1,$ and $4$ around the BAO scale of $r_1=100\Mpch$. In particular compare the $\ell=1$ panel here with that of Figure \ref{fig:patchy_w_bao_lowell} to aid in identifying the peak and trough the BAO induce in the 3PCF's dipole moment. The points in the peak are anti-correlated with those in the trough, as shown  in Figure \ref{fig:reduced_covar} (second tile on the diagonal).  These points are therefore more constraining than the error bars shown would suggest. The error bars plotted are the diagonal of the covariance matrix, and the $\chi^2/{\rm d.o.f.} = 107.64/107$.}
\label{fig:data_bao_lowell}
\end{figure}

\begin{figure}
\centering
\includegraphics[scale=1]{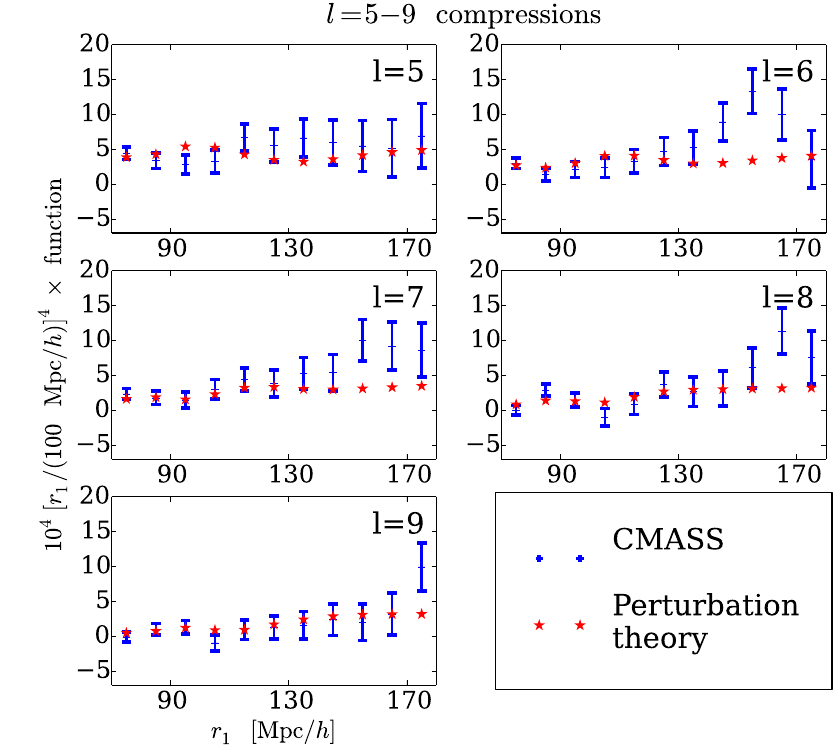}%updated. caption updated. 21 nov.
\caption{Same as Figure \ref{fig:data_bao_lowell} but now for multipoles $\ell=5-9$. These higher multipoles appear noisier than their lower-$\ell$ counterparts, as indicated by the larger number of points more than $1\sigma$ distant from the model. While the error bars are similar in magnitude to those in Figure \ref{fig:data_bao_lowell}, the signal is reduced relative to the largest in Figure \ref{fig:data_bao_lowell} (i.e. $\ell=2$ and $3$).}
\label{fig:data_bao_hiell}
\end{figure}

We fit the CMASS data in the same way as the mocks, with the results displayed Figures \ref{fig:data_bao_lowell} and \ref{fig:data_bao_hiell}. The error bars plotted are again from the diagonal of the covariance matrix. Again notice the match between data and PT. Though the match looks visually worse for the CMASS data than for the mocks, in fact the $\chi^2/{\rm d.o.f.} =107.64/107=1.01$, meaning the PT model is adequate for the dataset, whereas for the mocks it was not. 

In the case of the mocks, one might have worried that, since they are based on 2LPT, by construction PT would fit them well. However, the fair match between data and model shows that on these large scales and in our compressed basis, leading-order PT describes the 3PCF with reasonable fidelity.

The best-fit parameters for the data are in Table 1. We measure the redshift-space linear bias $b_1$ with $2.60\%$ precision at fixed $\sigma_8$. Further, in concert with the power spectrum or 2PCF, our work can be used to place a constraint on $\sigma_8$. If the power spectrum, $P\propto b_1^2\sigma_8^2$, and the redshift-space to real-space rescaling are known to high accuracy, then dividing $b_1^3\sigma_8^4$ by $P^{3/2}$ isolates $\sigma_8$ with $7.8\%$ precision (this approach was originally proposed in Fry 1994).

\begin{figure}
\centering
\includegraphics[scale=1]{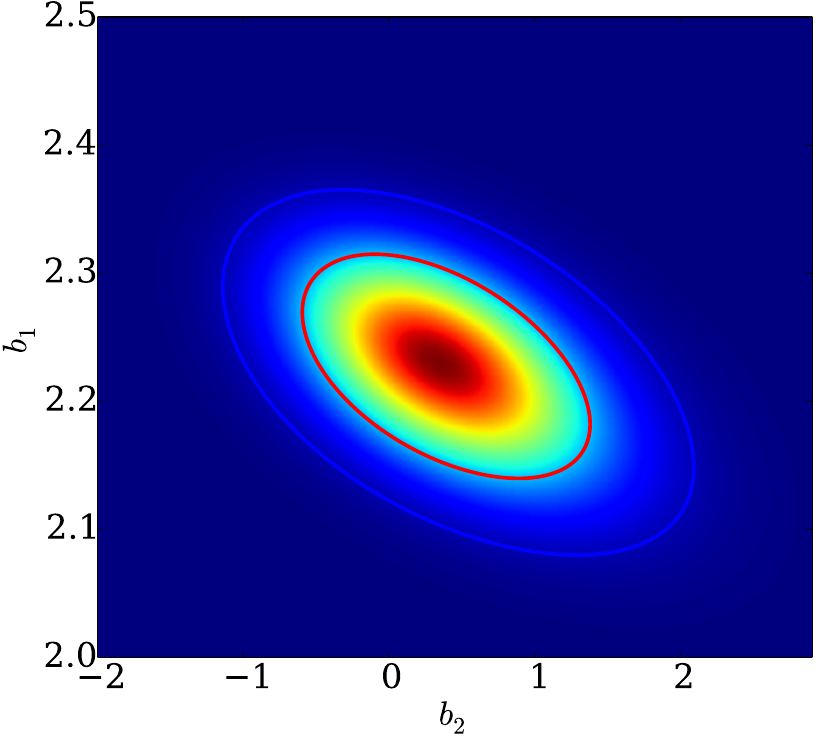}%updated.
\caption{The probability contours for the redshift-space biases $b_1$ and $b_2$ having marginalized over the integral constraint. The red ellipse contains $68\%$ of the probability and the light blue $95\%$. One can see that our measurement obtains a good constraint on $b_1$ but has very little constraining power on $b_2$, a conclusion borne out quantitatively by the large error bar on $b_2$ relative to that on $b_1$ quoted in Table 1.} 
\label{fig:data_b1_b2}
\end{figure}

We obtain essentially no constraint on $b_2$. On these large scales, the 3PCF in our compressed basis seems to be insensitive to the redshift-space non-linear bias aside from indicating its existence. Figure \ref{fig:data_b1_b2} shows the probability of a given $b_1$ and $b_2$ having marginalized over the integral constraint (encoded in $c$; see \S\ref{subsec:int_const}).  The elliptical appearance of the iso-probability contours means that $b_1$ and $b_2$ are roughly Gaussian-distributed. The ellipses drawn in red and light blue show $68\%$ and $95\%$ containment regions and do not assume Gaussianity; we simply integrate over the region until we reach these containments. The greater length of the ellipses in the $b_2$ direction illustrates that we do not obtain much constraint on $b_2$.  
%updated 21 nov.

%Fit to mocks using PT with BAO, ell = 0-4.

%Fit to mocks using PT with BAO, ell = 5-9.

%Fit to data using PT with BAO, ell=0-4.

%Fit to data using PT with BAO, ell=5-9.

%Confidence ellipse for b1-b2.

\subsection{Searching for the BAO}
\label{subsec:bao_search}
To determine the significance of a BAO signal in our compressed multipole measurements of the 3PCF, we fit PT predictions for the 3PCF computed using the no-wiggle model to both mocks and data, and compute the $\Delta \chi^2$ relative to our fits of \S\ref{subsec:md_results}, which used the physical power spectrum equation (\ref{eqn:p_template}). We emphasize that the BAO significance always stems from comparing the no-wiggle model to the physical power spectrum model. In the present work we do not fit for the BAO scale itself to extract distance information, but this will be a direction of future work.

\begin{figure}
\centering
\includegraphics[scale=1]{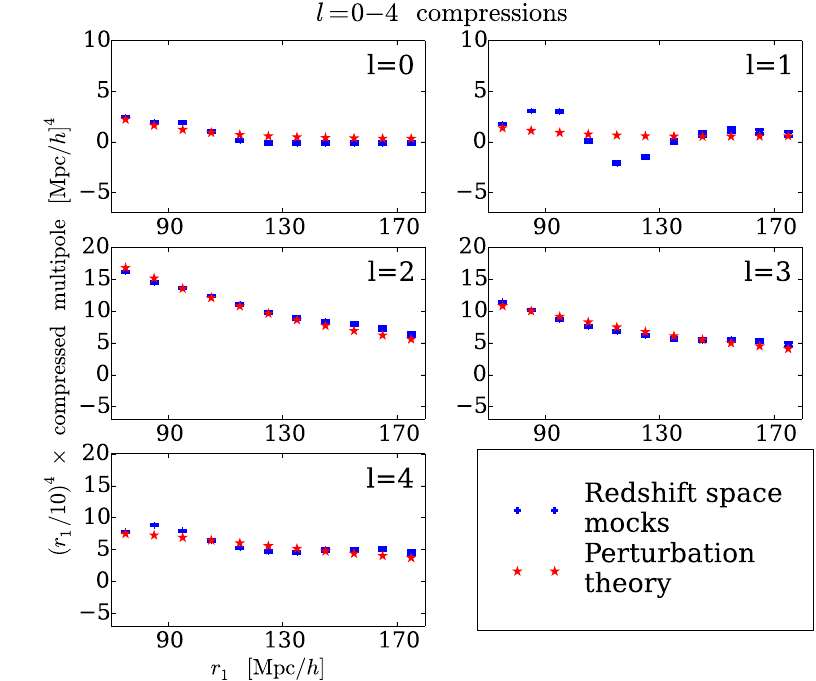}
\caption{A fit of PT predictions, computed using the no-wiggle power spectrum, to the compressed 3PCF's multipoles $\ell=0-4$ for the mean of 299 \textsc{MultiDark-Patchy} mocks. In general the PT with no-wiggle is not a bad fit except around the BAO scale of $r_1=100\Mpch$; this is especially so in $\ell=0, 1,$ and $4$.  This indicates a clear preference for a BAO feature in the 3PCF. The best-fit parameter values are $b_1 = 2.362\pm 0.003, \;b_2=0.16\pm 0.03,\;c=-0.0020 \pm 0.0006$ with $\chi^2/{\rm d.o.f} = 3639.88/107$. The error bars plotted are the diagonal of the covariance matrix scaled down appropriately for the mean of $299$ mocks.}%numbers updated: 7 dec.
\label{fig:patchy_no_bao}
\end{figure}

For the mean of the \textsc{MultiDark-Patchy} mocks, we find a clear preference for the BAO model. The comparison is between Figures \ref{fig:patchy_w_bao_lowell} (with BAO) and \ref{fig:patchy_no_bao} (without BAO). The $\Delta\chi^2$ is $3234.34$, meaning if we had a survey volume $299$ times as large as CMASS we would expect a $56.9\sigma$ BAO detection even in our compressed 3PCF. The reason for the large $\chi^2$ penalty of the no-BAO model over the model with BAO can be seen by visual comparison of Figures \ref{fig:patchy_w_bao_lowell} and \ref{fig:patchy_no_bao} around the BAO scale of $r_1=100\Mpch$, most prominently in $\ell=1$ but also in $\ell=0$ and $4$.

\begin{figure}
\centering
\includegraphics[scale=1]{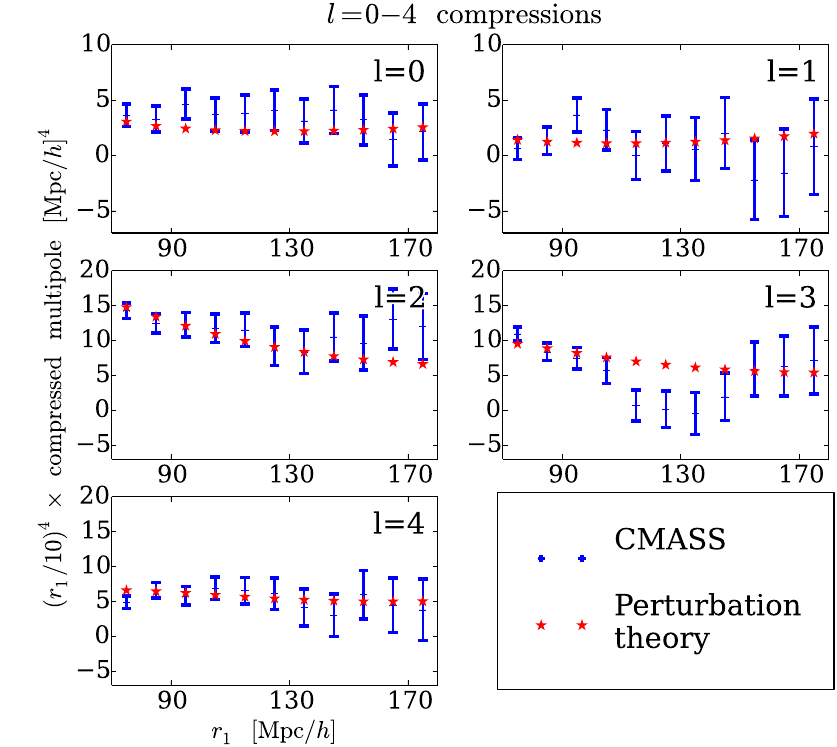}%updated. caption updated. 21 nov.
\caption{A fit of PT predictions, computed using the no-wiggle power spectrum, to the compressed 3PCF's multipoles $\ell=0-4$ for the CMASS sample. Notice the failure of the no-wiggle model especially in $l=0, 1,$ and $4$ at the BAO scale of $r_1=100\Mpch$. The parameter values are $b_1 =2.21\pm 0.06, \;b_2 =0.1\pm 0.6$ and $c=-0.022\pm 0.007$ and $\chi^2/{\rm d.o.f.} =115.22/107$. The error bars plotted are the diagonal of the covariance matrix.} %numbers updated: 7 dec.
\label{fig:data_no_bao}
\end{figure}

For the CMASS data, we again find a preference for the BAO, with $\Delta\chi^2=7.58$, meaning a  $2.8\sigma$ preference for the BAO.  One can see comparing Figures \ref{fig:data_bao_lowell} and \ref{fig:data_no_bao} that both physical and no-wiggle models fit the data reasonably well, but that around the BAO scale of $r_1=100\Mpch$ the no-wiggle model fits less well. Scaling the $56.9\sigma$ detection from the $299$ mocks' mean down by $\sqrt{299}$ to mirror the volume of CMASS, we expect on average a $3.29\sigma$ detection of the BAO, indicating our result from the CMASS data is plausible. We expect the BAO feature's significance to increase when we use the full multipole coefficients of the 3PCF rather than the compressions presented here, a direction of future work.
%updated 21 nov.

\begin{table}
\label{table:best_fit_parameters}
\begin{tabular}{|c|c|c|c|c|}
\hline 
\;\;\;\;\; & $b_1$ & $b_2$ & $c$& \specialcell{$\Delta\chi^2\;{\rm no\; BAO}$\\vs.\;BAO}\tabularnewline
\hline 
\specialcell{\textsc{MultiDark-}\\\textsc{Patchy} mocks} & \specialcell{$2.390$\\ $\pm0.003$} & \specialcell{$0.32$\\ $\pm0.04$} & \specialcell{$0.0000$\\$\pm0.0006$} &$3234.34$\tabularnewline
\hline %mock numbers updated: 7 december.
\specialcell{\textsc{CMASS} SDSS\\ DR12 sample} & \specialcell{$2.23$\\ $\pm0.06$} &\specialcell{$0.3$\\ $\pm0.7$} & \specialcell{$-0.023$\\ $\pm 0.007$} &$7.58$\tabularnewline
\hline %data numbers updated: 7 december.
\end{tabular}
\centering
\caption{Table of best-fit parameters for \textsc{MultiDark-Patchy} mocks and CMASS data.  The biases are redshift-space quantities, and $c$ encodes the integral constraint (\S\ref{subsec:int_const}).  The last column describes the $\chi^2$ penalty a no-BAO model (\S\ref{subsec:bao_search}) pays over a model with BAO.}
\end{table}

Finally, $b_2$ changes from physical to no-wiggle models for both the mocks and the data. This change likely occurs because the no-wiggle power spectrum differs slightly from the physical power spectrum on small scales and $b_2$ changes to absorb this difference.

\section{Conclusions}
\label{sec:conclusions}

We have measured the large-scale 3PCF of the SDSS DR12 CMASS sample of 777,202 LRGs. The novel multipole algorithm of SE15b permits us for the first time to take advantage of all triangle configurations. We have used a compression scheme first developed in SE15a to reduce the dimension of the full 3PCF multipole coefficients and to avoid the triangles for which PT is likely invalid.  We have shown that in this basis the analytic covariance matrix of SE15b, which assumes a Gaussian Random Field density and a boundary-free survey, is a good match to the empirical covariance matrix derived from 299 \textsc{MultiDark-Patchy} mock catalogs.  Using our analytic covariance matrix with volume and shot noise derived from the mocks, we have fit for the redshift-space linear and non-linear bias as well as a constant to marginalize over possible failure to satisfy the integral constraint.  We measure the redshift-space linear bias with $2.6\%$ precision.  {\bf We also find a $2.8\sigma$ preference for the BAO in the data} by comparing  a physical model with BAO to the Eisenstein \& Hu (1998) ``no-wiggle'' model.  

The present work uses the largest number of galaxies to date for the 3PCF by a factor of roughly 4, and is unique in exploiting all triangle configurations. We also measure the 3PCF on significantly larger scales (roughly a factor of 2) than previous works. Our error bar on the redshift-space linear bias is competitive with that found on the real-space linear bias from recent bispectrum studies such as Gil-Mar\'in et al. (2015), which measures the combination $b_1^{1.40}\sigma_8(z_{\rm eff})$ with $3.5\%$ precision (their Table 2). If $\sigma_8$ is fixed, as we do in the present work, their measurement translates to $2.6\%$ precision on $b_1$.

Gil-Mar\' in et al. (2015) use $690,827$ LRGs in the CMASS sample of DR11 for their bispectrum, finding $b_1^{1.40}\sigma_8(z_{\rm eff}) = 1.672\pm 0.060$. This work uses a detailed, complex model for removing RSD to obtain the real-space bias. Using our value of $\sigma_8$ to convert $b_1^{1.40}\sigma_8$ to $b_1$ translates this measurement to a real-space linear bias $b_1 = 2.03$. Gil-Mar\' in et al. (2015) also measure $b_1$ and $b_2$ from the power spectrum, finding $b_1 = 2.086$ and $b_2=0.902$; they do not quote errorbars on the parameters derived from the power spectrum. 

Our measured redshift-space $b_1$ is likely enhanced by $\sim 20-25\%$ relative to its real-space value based on the rescaling factors computed in \S\ref{subsec:rsd}.  Reducing our value of $2.23$ by these factors leads to an inferred real-space $b_1\sim 1.77-1.85$. However, we caution that the rescaling factors used here are estimates and so there is some uncertainty in how our redshift-space bias maps to a real-space bias. The sample of Gil-Mar\'in et al. (2015) is a proper subset of our own (though $\sim 90\%$), so one would not necessarily expect the bias values to be equal. Furthermore, the bias model of Gil-Mar\'in et al. (2015) is Lagrangian and includes a tidal tensor bias, while our model is Eulerian and does not include this bias. Consequently any quantitative comparison should be taken with considerable caution at present.

From the projected 2PCF, Guo et al. (2013) infer a real-space linear bias of $b_1=2.16\pm 0.01$ using subsamples from the CMASS sample of DR9 with a maximum of $260,000$ galaxies. Guo et al. (2013) uses projected separations of $3-25\Mpch$ and a detailed model of non-linear structure formation. That work's cosmology is WMAP7, with $\sigma_8(z=0) =0.8$, whereas in the present work $\sigma_8(z=0)=0.8288$. Rescaling the Guo et al. (2013) measurement to our $\sigma_8$ yields a real-space linear bias of $b_1=2.1$. Note that the projected measurement is insensitive to RSD. 

From the full 2PCF, Ross et al. (2014) uses $540,505$ CMASS galaxies in DR10 to measure the real-space linear bias, finding $b_1=1.96\pm 0.05$. $\sigma_8(z=0)=0.8$ for this work, and rescaling as discussed above reduces this value to $b_1=1.89$, comparable to our expected real-space range of $b_1\sim 1.77-1.85$. We caution that the galaxy samples used in Guo et al. (2013) and in Ross et al. (2014) are again proper subsets of our own (roughly $30\%$ and $70\%$, respectively) so one would not necessarily expect the bias values to be equal.

The 2PCF and 3PCF depend on different combinations of $b_1$ and $\sigma_8$ (Fry 1994), so measuring the two in concert can permit isolating a value for each. In the present work, we have not pursued a combined fit to the 2PCF and 3PCF; future work might explore this approach. Such work would require the covariance between the 2PCF and the 3PCF, a topic explored for the Fourier-space analogs in Sefusatti et al. (2006), who find that squeezed triangles dominate this covariance. As our compression scheme avoids such triangles, it may be beneficial in the context of joint 2PCF-3PCF fitting. 

Regarding squeezed triangles, the position-dependent 2-point correlation function, developed in Chiang et al. (2015), likely offers interesting complementary information not captured by our compression. This technique evaluates the 2PCF within sub-volumes conditioned on the mean overdensity of those subvolumes, and is therefore sensitive to an integral of the 3PCF dominated by squeezed triangles.  Chiang et al. (2015) have already applied the technique to the SDSS DR10 CMASS sample to measure the non-linear bias $b_2$. In future it would be worthwhile to explore joint parameter constraints using the position-dependent 2PCF and our compressed 3PCF.

We will also explore more complicated bias models, such as Lagrangian schemes (shown to be more accurate in Chan et al. 2012) that include tidal tensor bias (McDonald \& Roy 2009; Baldauf et al. 2012).  Adding a tidal tensor bias might modestly raise our error bars on the measured biases. However it is unlikely to alter the BAO significance because the tidal tensor enters at $\ell=0$ and $2$ pre-cyclically whereas the BAO's dominant contribution is at $\ell =1$ (see \S\ref{sec:model}). Additional modeling of RSD to permit precise extraction of the real-space bias values will also be important moving forward. It may also be worthwhile to further explore the impact of observational systematics on the 3PCF, for instance by reanalyzing the sample with systematic weights such as the stellar density weight turned off.

Finally, in future work we will extend our BAO search from the compressed statistic to the full multipole moments of the 3PCF. We developed the compressed approach to ease handling of the covariance matrix and fitting as well as to avoid regimes where PT is invalid, but it may be overly conservative. In practice we have found that the covariance matrix and fitting are computationally by far not the limiting steps in our analysis, and the good agreemeent between the compressed PT and our measurements argues that there is some room to incorporate more strongly squeezed triangles before PT breaks down.

Regarding the BAO, Gazta\~naga et al. (2009)'s finding of a $2-3\sigma$ preference using a single triangle configuration measured from $40,000$ LRGs suggests that our measurement of the 3PCF might contain an even more significant BAO feature than we have found. Gazta\~naga et al. (2009) did find a high value of the baryon fraction $\fb = 0.28$, significantly larger than the Planck-compatible value $\fb = 0.16$ used in the present work; this is likely related to the strength of their BAO feature. Since our work uses many triangle configurations but also compresses, the significance of their BAO evidence cannot be simply scaled up as the square root of survey volume or galaxy number.  Nonetheless, this is an additional motivation for future work examining analysis variations in the context of the BAO. 

\section*{Acknowledgments}

ZS thanks Chia-Hsun Chuang for substantial contributions to this work. ZS further thanks Neta Bahcall, Patricia Burchat, Blakesley Burkhart, Aaron Bray, Douglas Finkbeiner, Lehman Garrison, Margaret Geller, JR Gott III, James Guillochon, Elizabeth Krause, Abraham Loeb, Ramesh Narayan, Stephen Portillo, Kate Rubin, Jeff Scargle, Marcel Schmittfull, Uro\v s Seljak, Joshua Suresh, David Spergel, Alexander Wiegand, Risa Wechsler, and Matias Zaldarriaga for valuable discussions. This material is based upon work supported by the National Science Foundation Graduate Research Fellowship under Grant No. DGE-1144152; DJE is supported by grant DE-SC0013718 from the U.S. Department of Energy. AJC is supported by supported by the European Research Council under the European Community's Seventh Framework Programme FP7-IDEAS-Phys.LSS 240117. Funding for this work was partially provided by the Spanish MINECO under projects FPA2011-29678-C02-02 and MDM-2014-0369 of ICCUB (Unidad de Excelencia 'Mar{\'\i}a de Maeztu'). HGM acknowledges support from the Labex ILP (reference ANR-10-LABX-63), part of the Idex SUPER, receiving state financial aid managed by the Agence Nationale de la Recherche, as part of the programme Investissements d'avenir under the reference ANR-11-IDEX-0004-02. FSK acknowledges support from the Leibniz Society for the Karl-Schwarzschild fellowship.

Funding for SDSS-III has been provided by the Alfred P. Sloan Foundation, the Participating Institutions, the National Science Foundation, and the U.S. Department of Energy Office of Science. The SDSS-III web site is http://www.sdss3.org/.
SDSS-III is managed by the Astrophysical Research Consortium for the Participating Institutions of the SDSS-III Collaboration including the University of Arizona, the Brazilian Participation Group, Brookhaven National Laboratory, University of Cambridge, Carnegie Mellon University, University of Florida, the French Participation Group, the German Participation Group, Harvard University, the Instituto de Astrofisica de Canarias, the Michigan State/Notre Dame/JINA Participation Group, Johns Hopkins University, Lawrence Berkeley National Laboratory, Max Planck Institute for Astrophysics, Max Planck Institute for Extraterrestrial Physics, New Mexico State University, New York University, Ohio State University, Pennsylvania State University, University of Portsmouth, Princeton University, the Spanish Participation Group, University of Tokyo, University of Utah, Vanderbilt University, University of Virginia, University of Washington, and Yale University.

\section*{References}
\hangindent=1.5em
\hangafter=1
\noindent Abazajian K. N. et al., 2009, ApJS, 182, 543.

\hangindent=1.5em
\hangafter=1
\noindent Aihara H. et al., 2011, ApJS, 193, 29.

\noindent Alam S. et al., 2015, preprint (arXiv:1501.00963). %SDSS DR11 and DR12 paper. http://adsabs.harvard.edu/cgi-bin/bib_query?arXiv:1501.00963
%done.

\hangindent=1.5em
\hangafter=1
\noindent Anderson et al. 2012, MNRAS 427, 4, 3435-3467.%http://adsabs.harvard.edu/abs/2012MNRAS.427.3435A

\hangindent=1.5em
\hangafter=1
\noindent Anderson et al. 2014, MNRAS 441, 1, 24-62.%http://adsabs.harvard.edu/abs/2014MNRAS.441...24A

\hangindent=1.5em
\hangafter=1
\noindent Baldauf T., Seljak U., Desjacques V. \& McDonald P., 2012, PRD 86, 8.%http://adsabs.harvard.edu/cgi-bin/bib_query?arXiv:1201.4827
%uros tidal tensor cite.

\hangindent=1.5em
\hangafter=1
\noindent Bardeen J.M., Steinhardt P.J. \& Turner M.S., 1983, PRD, 28, 4, 679-693.%http://adsabs.harvard.edu/abs/1983PhRvD..28..679B
%Spontaneous creation of almost scale-free density perturbations in an inflationary universe

\hangindent=1.5em
\hangafter=1
\noindent Bernardeau F., Colombi S., Gazta\~{n}aga E., Scoccimarro R., 2002, Phys. Rep., 367, 1.

\hangindent=1.5em
\hangafter=1
\noindent Blake C. \& Glazebrook K., 2003, ApJ 594, 2, 665-673.%http://adsabs.harvard.edu/abs/2003ApJ...594..665B

\hangindent=1.5em
\hangafter=1
\noindent Blanton M. et al., 2003, AJ, 125, 2276.

\hangindent=1.5em
\hangafter=1
\noindent Bolton A. et al., 2012, AJ, 144, 144.

\hangindent=1.5em
\hangafter=1
\noindent Burden A., Percival W.J., Manera M., Cuesta A.J., Vargas-Maga\~na M. \& Ho S., 2014, MNRAS, 445, 3, 3152-3168.%http://adsabs.harvard.edu/abs/2014MNRAS.445.3152B

\hangindent=1.5em
\hangafter=1
\noindent Chan K.C., Scoccimarro R. \& Sheth R.K., 2012, PRD 85, 8, 083509.%http://adsabs.harvard.edu/cgi-bin/bib_query?arXiv:1201.3614
%Lagrangian biasing.

\hangindent=1.5em
\hangafter=1
\noindent  Chiang C.-T., Wagner C., S\'anchez A.G., Schmidt F. \& Komatsu E., 2015, JCAP 09, 28.%http://adsabs.harvard.edu/abs/2015JCAP...09..028C

\hangindent=1.5em
\hangafter=1
\noindent Chuang C.-H. et al., 2015, MNRAS 452, 1, 686-700.%http://adsabs.harvard.edu/abs/2015MNRAS.452..686C
%nIFTy comparison of clustering for mocks and full N-body.

\hangindent=1.5em
\hangafter=1
\noindent Cuesta A.J. et al., 2015, preprint (arXiv:1509.06371).%http://adsabs.harvard.edu/cgi-bin/bib_query?arXiv:1509.06371

\hangindent=1.5em
\hangafter=1
\noindent Dawson K.S. et al., 2013, AJ, 145, 10.

\hangindent=1.5em
\hangafter=1
\noindent Dekel A. \& Rees M.J., 1987, Nature 326:455-462.%review of evidence for biased galaxy formation; early reference.

\hangindent=1.5em
\hangafter=1
\noindent Desjacques V. \& Seljak U., 2010, Class. and Quant. Grav. 27, 12.

\hangindent=1.5em
\hangafter=1
\noindent  Doi M. et al., 2010, AJ, 139, 1628.

\hangindent=1.5em
\hangafter=1
\noindent Eisenstein D.J. \& Hu W., 1998, ApJ 496, 605.

\hangindent=1.5em
\hangafter=1
\noindent Eisenstein D.J., Hu W. \& Tegmark M., 1998, ApJ  504:L57-L60.%http://background.uchicago.edu/~whu/Papers/hubble.pdf

\hangindent=1.5em
\hangafter=1
\noindent Eisenstein D.J., Seo H.-J., Sirko E. \& Spergel D., 2007, ApJ 664, 2, 675-679.%http://adsabs.harvard.edu/abs/2007ApJ...664..675E

\hangindent=1.5em
\hangafter=1
\noindent Eisenstein D.J., Seo H.-J. \& White M., 2007, ApJ 664, 2, 660-674.%http://adsabs.harvard.edu/abs/2007ApJ...664..660E

\hangindent=1.5em
\hangafter=1
\noindent Eisenstein D.J. et al., 2011, AJ, 142, 72.

%turn on if we mention FKP weights.
\iffalse
\hangindent=1.5em
\hangafter=1
\noindent Feldman H.A., Kaiser N. \& Peacock J.A., 1994, ApJ, 426, 23.
\fi

\hangindent=1.5em
\hangafter=1
\noindent Feldman H.A., Frieman J.A., Fry J.N. \& Scoccimarro R., 2001, PRL 86, 1434.

\hangindent=1.5em
\hangafter=1
\noindent Fry J.N. \& Seldner M., 1982, ApJ, 259, 474.

\hangindent=1.5em
\hangafter=1
\noindent Fry J.N., 1984, ApJ 279, 499-510.%http://adsabs.harvard.edu/abs/1984ApJ...279..499F

\hangindent=1.5em
\hangafter=1
\noindent Fry J.N. \& Gazta\~naga E., 1993, ApJ 413, 2, 447-452.%http://adsabs.harvard.edu/abs/1993ApJ...413..447F
%biasing, Roman cite I think.

\hangindent=1.5em
\hangafter=1
\noindent Fry J.N., 1994, PRL 73, 215.%Disentangling b1/sigma8.

\hangindent=1.5em
\hangafter=1
\noindent Fukugita M. et al., 1996, AJ, 111, 1748.

\hangindent=1.5em
\hangafter=1
\noindent Gazta\~naga E. \& Scoccimarro R., 2005, MNRAS, 361, 824.

\hangindent=1.5em
\hangafter=1
\noindent  Gazta\~naga E., Cabr\'e A., Castander F., Crocce M. \& Fosalba P., 2009, MNRAS 399, 2, 801-811.%http://adsabs.harvard.edu/cgi-bin/bib_query?arXiv:0807.2448
%BAO in the 3PCF.

\hangindent=1.5em
\hangafter=1
\noindent Giannantonio T. et al., 2014, PRD 89, 2, 023511.%http://adsabs.harvard.edu/cgi-bin/bib_query?arXiv:1303.1349
%LSS constraints on PNG.

\hangindent=1.5em
\hangafter=1
\noindent Gil-Mar\'in H., Nore\~na J., Verde L., Percival W.J., Wagner C., Manera M. \& Schneider D.P., 2015, MNRAS 451, 1, 539-580.%http://adsabs.harvard.edu/cgi-bin/bib_query?arXiv:1407.5668

\hangindent=1.5em
\hangafter=1
\noindent Gunn J.E. et al., 1998, AJ, 116, 3040.

\hangindent=1.5em
\hangafter=1
\noindent Gunn J.E. et al., 2006, AJ, 131, 2332.

\hangindent=1.5em
\hangafter=1
\noindent Guo H. et al., 2013, ApJ 767, 2, 122.%http://adsabs.harvard.edu/cgi-bin/bib_query?arXiv:1212.1211
%projected 2PCF linear bias measurement; 2.16.

\hangindent=1.5em
\hangafter=1
\noindent Guo H. et al., 2014, ApJ 780, 2, 139.%http://adsabs.harvard.edu/cgi-bin/bib_query?arXiv:1303.2609
%3PCF up to 40 Mpc/h with SDSS DR7.

\hangindent=1.5em
\hangafter=1
\noindent Guo H. et al., 2015, MNRAS 449, 1, L95-L99.%http://adsabs.harvard.edu/cgi-bin/bib_query?arXiv:1409.7389

\hangindent=1.5em
\hangafter=1
\noindent Groth E. J. \& Peebles P. J. E., 1977, 217, 385.

\hangindent=1.5em
\hangafter=1
\noindent Hamilton A.J.S, 1998, in ``The Evolving Universe: Selected Topics on Large-Scale Structure and on the Properties of Galaxies,'' Dordrecht: Kluwer.%http://adsabs.harvard.edu/cgi-bin/bib_query?arXiv:astro-ph/9708102
%done

\hangindent=1.5em
\hangafter=1
\noindent Hu W. \& Haiman Z., 2003, PRD 68, 6, 063004.%http://adsabs.harvard.edu/abs/2003PhRvD..68f3004H

\hangindent=1.5em
\hangafter=1
\noindent Jackson J.C., 1972, MNRAS 156, 1.%http://adsabs.harvard.edu/cgi-bin/bib_query?arXiv:0810.3908

\hangindent=1.5em
\hangafter=1
\noindent Jing Y.P. \& B\"{o}rner G., 2004, ApJ 607, 140.

\hangindent=1.5em
\hangafter=1
\noindent Kaiser N. 1984, ApJL 284, L9-12.

\hangindent=1.5em
\hangafter=1
\noindent Kaiser N., 1987, MNRAS 227, 1.

\hangindent=1.5em
\hangafter=1
\noindent Kitaura F.-S. \& He{\ss} S., 2013, MNRAS 435, 1, L78-L82.%http://adsabs.harvard.edu/cgi-bin/bib_query?arXiv:1212.3514

\hangindent=1.5em
\hangafter=1
\noindent Kitaura F.-S., Yepes G. \& Prada F., 2014, MNRAS 439, L21.

\hangindent=1.5em
\hangafter=1
\noindent Kitaura F.-S., Gil-Mar\'in H., Sc\'occola C.G., Chuang C.-H., M\"{u}ller V., Yepes G. \& Prada F., 2015a, MNRAS 450, 2, 1836-1845.%http://adsabs.harvard.edu/abs/2015MNRAS.450.1836K

\hangindent=1.5em
\hangafter=1
\noindent Kitaura F.-S. et al., 2015b, preprint (arXiv:1509.06400).%http://adsabs.harvard.edu/cgi-bin/bib_query?arXiv:1509.06400

\hangindent=1.5em
\hangafter=1
\noindent Kulkarni G., Nichol R., Sheth R., Seo H.-J., Eisenstein D.J. \& Gray A., 2007, MNRAS 378, 3, 1196-1206.%http://adsabs.harvard.edu/abs/2007MNRAS.378.1196K
%DR3, 50967 LRGs, three scales, 4, 7, and 10 mpc/h. b1 = 1.87 \pm 0.07.

\hangindent=1.5em
\hangafter=1
\noindent Lewis A., 2000, ApJ, 538, 473.

\hangindent=1.5em
\hangafter=1
\noindent Linder E.V., 2003, PRD 68, 8, 083504.%http://adsabs.harvard.edu/abs/2003PhRvD..68h3504L

\hangindent=1.5em
\hangafter=1
\noindent Lupton R., Gunn J.E., Ivezi\'c Z., Knapp G. \& Kent S., 2001, ``Astronomical Data Analysis Software and Systems X'', v. 238, 269.

\hangindent=1.5em
\hangafter=1
\noindent Mar\'in F.A., 2011, ApJ 737, 2, 97.%http://adsabs.harvard.edu/cgi-bin/bib_query?arXiv:1011.4530
%Uses SDSS DR7 out to 90 Mpc/h. Focus on bias and sigma8.

\hangindent=1.5em
\hangafter=1
\noindent Mar\'in F.A. et al., 2013, MNRAS 432, 4 2654.%http://adsabs.harvard.edu/cgi-bin/bib_query?arXiv:1303.6644
%187,000 galaxies in WiggleZ; constrained sigma 8 by using sims to get biases first and then marginalizing.

\hangindent=1.5em
\hangafter=1
\noindent McBride C., Connolly A.J., Gardner J.P., Scranton R., Newman J., Scoccimarro R., Zehavi I., Schneider D.P., 2011a, ApJ, 726, 13.
%http://adsabs.harvard.edu/abs/2011ApJ...726...13M
%done

\hangindent=1.5em
\hangafter=1
\noindent McBride K., Connolly A.J., Gardner J.P., Scranton R., Scoccimarro R., Berlind A., Marin F., Schneider D.P., 2011b, ApJ, 739, 85
%http://adsabs.harvard.edu/abs/2011ApJ...739...85M
%done

\hangindent=1.5em
\hangafter=1
\noindent McDonald P. \& Roy A., 2009, JCAP 0908, 020.%Tidal tensor bias reference.
%arXiv: 0902.0991

\hangindent=1.5em
\hangafter=1
\noindent Nichol R.C. et al., 2006, MNRAS 368, 4, 1507-1514.%http://adsabs.harvard.edu/abs/2006MNRAS.368.1507N
%36,738 LRGs SDSS; not sure what DR. Maximal scale >~10 mpc/h?

\hangindent=1.5em
\hangafter=1
\noindent Osumi K., Ho S., Eisenstein D.J. \& Vargas-Maga\~na M., 2015, preprint (arXiv:1505.00782).%http://adsabs.harvard.edu/cgi-bin/bib_query?arXiv:1505.00782
%BAO constant fit robustness.

\hangindent=1.5em
\hangafter=1
\noindent Padmanabhan N. et al., 2008, ApJ, 674, 1217.

\hangindent=1.5em
\hangafter=1
\noindent Padmanabhan N., White M. \& Cohn J.D., 2009, PRD 79, 6, 063523.%http://adsabs.harvard.edu/abs/2009PhRvD..79f3523P

\hangindent=1.5em
\hangafter=1
\noindent Pan J. \& Szapudi I., 2005, MNRAS 362, 4, 1363-1370.%http://adsabs.harvard.edu/cgi-bin/bib_query?arXiv:astro-ph/0505422

\hangindent=1.5em
\hangafter=1
\noindent Peebles P.J.E. \& Groth E.J., 1975, ApJ, 196, 1.

\hangindent=1.5em
\hangafter=1
\noindent Peebles P.J.E., 1980, The Large-Scale Structure of the Universe: Princeton University Press, Princeton.

\hangindent=1.5em
\hangafter=1
\noindent Percival W.J. et al., 2014, MNRAS 439, 3, 2531-2541.%http://adsabs.harvard.edu/abs/2014MNRAS.439.2531P

\hangindent=1.5em
\hangafter=1
\noindent Pier J.R. et al., 2003, AJ, 125, 1559.

\hangindent=1.5em
\hangafter=1
\noindent Planck Collaboration, Paper XIII, 2015, preprint (arXiv:1502.01589).%http://adsabs.harvard.edu/cgi-bin/bib_query?arXiv:1502.01589

\hangindent=1.5em
\hangafter=1
\noindent Planck Collaboration, Paper XVII, 2015, preprint (arXiv:1502.01592).%http://adsabs.harvard.edu/cgi-bin/bib_query?arXiv:1502.01592

\hangindent=1.5em
\hangafter=1
\noindent Reid B. et al., 2016, MNRAS 455, 2, 1553-1573.%http://adsabs.harvard.edu/cgi-bin/bib_query?arXiv:1509.06529

\hangindent=1.5em
\hangafter=1
\noindent Rodr\'iguez-Torres S. et al., 2015, preprint (arXiv:1509.06404).%http://adsabs.harvard.edu/cgi-bin/bib_query?arXiv:1509.06404

\hangindent=1.5em
\hangafter=1
\noindent Ross A.J. et al., 2012, MNRAS 424, 1, 564-590.%http://adsabs.harvard.edu/abs/2012MNRAS.424..564R
%Paper including constant term in BAO fits.

\hangindent=1.5em
\hangafter=1
\noindent Ross A.J. et al., 2013, MNRAS 428, 2, 1116-1127.%http://adsabs.harvard.edu/abs/2013MNRAS.428.1116R
%BOSS DR9 PNG constraints.

\hangindent=1.5em
\hangafter=1
\noindent Ross A.J. et al., 2014, MNRAS 437, 2, 1109-1126.%http://adsabs.harvard.edu/cgi-bin/bib_query?arXiv:1310.1106
%2PCF bias measurement with DR10, CMASS, 540,505 galaxies.

\hangindent=1.5em
\hangafter=1
\noindent Ross A.J. et al., 2015, in prep.

\hangindent=1.5em
\hangafter=1
\noindent Schmittfull M.M., Baldauf T. \& Seljak U, 2015, PRD 91, 4, 043530.%http://adsabs.harvard.edu/abs/2015PhRvD..91d3530S

\hangindent=1.5em
\hangafter=1
\noindent Schmittfull M.M., Feng Y., Beutler F., Sherwin B. \& Yat Chu, M., 2015, preprint (arXiv:1508.06972).%http://adsabs.harvard.edu/cgi-bin/bib_query?arXiv:1508.06972
%BAO reconstruction, how 3PCF enters.

\hangindent=1.5em
\hangafter=1
\noindent Schmittfull M.M.,  Regan D.M. \& Shellard E.P.S., 2013, PRD, 88, 6, 063512.%http://adsabs.harvard.edu/cgi-bin/bib_query?arXiv:1207.5678

\hangindent=1.5em
\hangafter=1
\noindent Scoccimarro R., 2000, ApJ 544, 593.%http://iopscience.iop.org/article/10.1086/317248
%ell=0 bispectrum FT method.

\hangindent=1.5em
\hangafter=1
\noindent Scoccimarro R., Couchman H.M.P. \& Frieman J.A., 1999, ApJ 517:531-540.%http://iopscience.iop.org/article/10.1086/307220/fulltext/

\hangindent=1.5em
\hangafter=1
\noindent Scoccimarro R., Feldman H., Fry J.N. \& Frieman J.A., 2001, ApJ 546, 2, 652-664.%http://adsabs.harvard.edu/abs/2001ApJ...546..652S

\hangindent=1.5em
\hangafter=1
\noindent Scoccimarro R., 2015, preprint (arXiv:1506.02729).%http://adsabs.harvard.edu/cgi-bin/bib_query?arXiv:1506.02729

\hangindent=1.5em
\hangafter=1
\noindent Sefusatti E., Crocce M., Pueblas S. \& Scoccimarro R., 2006, PRD 74, 2, 023522.%http://adsabs.harvard.edu/cgi-bin/bib_query?arXiv:astro-ph/0604505
%covariance of power spectrum and bispectrum.

\hangindent=1.5em
\hangafter=1
\noindent Sefusatti E. \& Komatsu E., 2007, PRD 76, 8, 083004.%http://adsabs.harvard.edu/abs/2007PhRvD..76h3004S
%Forecasts for PNG and bispectrum.

\noindent Seo H.J. \& Eisenstein D.J., 2003. ApJ 598, 2, 720-740.%http://adsabs.harvard.edu/abs/2003ApJ...598..720S

\hangindent=1em
\hangafter=1
\noindent Slepian Z. \& Eisenstein D.J., 2015a, MNRAS 448, 1, 9-26.%http://adsabs.harvard.edu/cgi-bin/bib_query?arXiv:1411.4052

\hangindent=1em
\hangafter=1
\noindent Slepian Z. \& Eisenstein D.J., 2015b, MNRAS 454, 4, 4142-4158.%http://adsabs.harvard.edu/cgi-bin/bib_query?arXiv:1506.02040

\hangindent=1em
\hangafter=1
\noindent Slepian Z. \& Eisenstein D.J., 2015c, MNRASL 455, 1, L31-L35.

\hangindent=1em
\hangafter=1
\noindent Slepian Z. \& Eisenstein D.J., 2015d, preprint (arXiv:1509.08199).%Transfer function paper. Change to MNRAS in press soon hopefully!

\hangindent=1em
\hangafter=1
\noindent  Smee S. et al., 2013, AJ, 146, 32.

\hangindent=1em
\hangafter=1
\noindent Smith J.A. et al., 2002, AJ, 123, 2121.

\hangindent=1em
\hangafter=1
\noindent Starobinsky A.A., 1982, Phys. Lett. B, 117, 3-4, 175-178.%http://adsabs.harvard.edu/abs/1982PhLB..117..175S
%Dynamics of phase transition in the new inflationary universe scenario and generation of perturbations.

\hangindent=1.5em
\hangafter=1
\noindent Szapudi I., Szalay A., 1998, ApJ, 494, L41.

\hangindent=1.5em
\hangafter=1
\noindent Szapudi I., 2004, ApJ, 605, L89.

\hangindent=1.5em
\hangafter=1
\noindent Verde L. et al., 2002, MNRAS, 335, 432.

\hangindent=1.5em
\hangafter=1
\noindent Wang Y., Yang X., Mo H.J., van den Bosch F.C. \& Chu Y., 2004, MNRAS, 353, 287.

\hangindent=1.5em
\hangafter=1
\noindent Xia J.-Q. , Baccigalupi C., Matarrese S., Verde L. \& Viel M., 2011, JCAP 8, 33.%arXiv:1104.5015.

\hangindent=1.5em
\hangafter=1
\noindent Xia J.-Q., Bonaldi A., Baccigalupi C.,  De Zotti G.,
Matarrese S., Verde L. \& Viel M., 2010, JCAP 8, 13.%http://arxiv.org/abs/1007.1969
%LSS fnl constraints.

\hangindent=1.5em
\hangafter=1
\noindent Xu X., Padmanabhan N., Eisenstein D.J., Mehta K.T. \& Cuesta A.J., 2012, MNRAS 427, 3, 2146-2167.%http://adsabs.harvard.edu/abs/2012MNRAS.427.2146X

\hangindent=1.5em
\hangafter=1
\noindent York D.G. et al., 2000, AJ, 120, 1579.

\end{document}